\documentstyle[prd,aps,floats]{revtex}
\begin{document}
\draft
%
%
\input epsf
\renewcommand{\topfraction}{0.8}
\preprint{CERN-TH/97-296, astro-ph/9711214}
\title{Quasi-open Inflation} 
\author{Juan Garcia-Bellido}
\address{TH-Division, C.E.R.N., CH-1211 Gen{\`e}ve 23, Switzerland}
\author{Jaume Garriga and Xavier Montes} 
\address{IFAE, Edifici C,
  Universitat Aut{\`o}noma de Barcelona, E-08193 Bellaterra, Spain}
\date{November 3, 1997} 
\maketitle

\begin{abstract}
  
We show that a large class of two-field models of single-bubble open
inflation do not lead to infinite open universes, as it was previously
thought, but to an ensemble of very large but finite inflating
`islands'. The reason is that the quantum tunneling responsible for
the nucleation of the bubble does not occur simultaneously along both
field directions and equal-time hypersurfaces in the open universe are
not synchronized with equal-density or fixed-field hypersurfaces. The
most probable tunneling trajectory corresponds to a zero value of the
inflaton field; large values, necessary for the second period of
inflation inside the bubble, only arise as localized fluctuations. The
interior of each nucleated bubble will contain an infinite number of
such inflating regions of comoving size of order $\gamma^{-1}$, where
$\gamma$ depends on the parameters of the model. Each one of these
islands will be a quasi-open universe. Since the volume of the
hyperboloid is infinite, inflating islands with all possible values of
the field at their center will be realized inside of a single
bubble. We may happen to live in one of those patches of comoving size
$d\lesssim \gamma^{-1}$, where the universe appears to be open. In
particular, we consider the ``supernatural'' model proposed by Linde
and Mezhlumian. There, an approximate $U(1)$ symmetry is broken by a
tunneling field in a first order phase transition, and slow-roll
inflation inside the nucleated bubble is driven by the
pseudo-Goldstone field. We find that the excitations of the
pseudo-Goldstone produced by the nucleation and subsequent expansion
of the bubble place severe constraints on this model. We also discuss
the coupled and uncoupled two-field models.  

\end{abstract}

\pacs{PACS numbers: 98.80.Cq \hspace{3.0cm} Preprint CERN-TH/97-296, 
\ astro-ph/9711214}


\section{Introduction}

In models of open inflation, which lead to a density parameter
$\Omega_0<1$, the ``horizon'' and ``flatness'' problems are solved by
two very different mechanisms. Although open inflation can be realized
with a single scalar, realistic models look more natural when the task
of solving each one of these two problems is entrusted to a different
scalar field.  Nevertheless, models with two fields introduce a host
of new effects which should be carefully investigated.  In particular,
as we shall see in this paper, most of the two-field models that have
been recently proposed do not give rise to an infinite open universe
but to a large inflating island of finite size: a quasi-open universe.

\begin{figure}[t]
\centering
\hspace*{-4mm}
\leavevmode\epsfysize=5.5cm \epsfbox{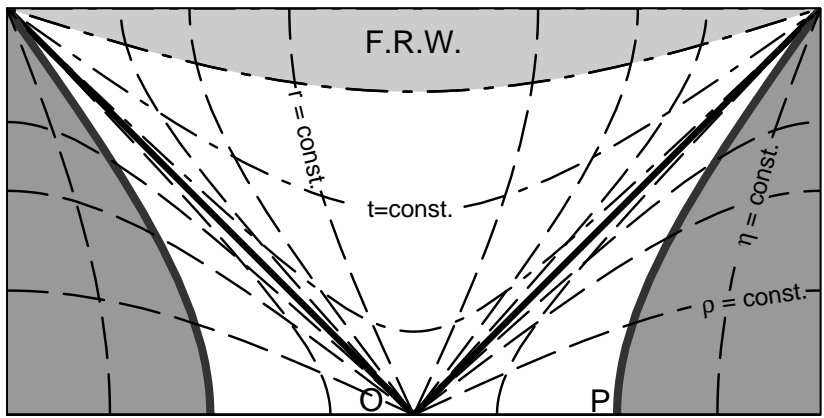}\\[3mm]
\caption[fig1]{\label{fig1} Conformal diagram showing a bubble expanding 
  in a de Sitter background. The bubble wall is represented as the
  thick grey line starting at the point P. It expands with constant
  acceleration along a $\eta = const.$ surface. The FRW open universe
  is inside the future light cone from the point O, which is the
  center of symmetry of the bubble solution.}
\end{figure}

The picture of open inflation is the following.  The universe starts
in a de Sitter phase driven by the potential energy of a scalar field
$\sigma$ which is trapped in a false vaccuum. This false vacuum decays
through quantum tunneling, and spherical bubbles of true vacuum
nucleate in the smooth de Sitter background. After nucleation, the
bubbles expand with constant acceleration, following a ``trajectory''
which is invariant under Lorentz transformations $O(3,1)$, see
Refs.~\cite{coleman,CdL} and Fig.~\ref{fig1}. Since this is also the
symmetry of an open Friedmann-Robertson-Walker (FRW) universe, the
$\sigma=const.$ surfaces in the interior of the bubble can be
identified with the $t=const.$ sections of an open universe
\cite{modelsopen1}.  In this way the symmetry of the bubble takes care
of the ``homogeneity'' problem. A second period of `slow-roll'
inflation inside the bubble, lasting for approximately 60 $e$-foldings,
would solve the ``flatness''
problem~\cite{modelsopen2,LM,induced,GBLhybrid}.

In open (and in quasi-open) inflation, the dynamics of bubble
nucleation and subsequent expansion turns out to be very important in
determining the spectrum of gravity waves and density perturbations.
The reason is that, unlike the case of stardard inflation, the amount
of slow-roll inflation is minimal and the ``initial conditions'' right
after the bubble nucleates are not washed out completely.  Thus, for
instance, quantum fluctuations of the slow-roll field generated
outside the bubble can penetrate to the interior~\cite{misao}, causing
perturbations whose wavelength is larger than the curvature scale.
These are the so-called supercurvature modes~\cite{super}. Also, the
scattering of tensor modes off the bubble wall determines the spectrum
of very long wavelength gravitational waves~\cite{tensor,wall1}. In the
limit of a weakly gravitating wall, this effect can be alternatively
described as a fluctuation of the bubble wall itself, which induces
supercurvature anisotropies inside the 
bubble~\cite{s1garcia,JG,s1misao,JDC}.

In principle, tunneling and slow-roll can be done by the same scalar
field~\cite{modelsopen2}, but this requires a very special form of the
inflaton potential $V$. Denoting by $H\equiv (8\pi G V/3)^{1/2}$ the
Hubble rate during inflation, a sharp barrier where $V''\gg H^2$ is
necessary\footnote{In the case $V''\ll H^2$ the phase transition can
  proceed via the Hawking-Moss instanton~\cite{HM}. However, this
  channel represents tunneling to the top of the barrier of a region
  of size $H^{-1}$, and does not lead to an open universe.} for bubble
nucleation~\cite{rama}. But this barrier must be right next to a
flatter region where $V''\ll H^2$, which is needed to make slow-roll
possible. Moreover, the duration of slow-roll, which depends on the
length of the plateau in the inflaton potential, has to be fine-tuned
to some extent, because a few $e$-foldings more or less can make the
difference between an almost flat universe and an almost empty one.

As mentioned above, models with two fields were introduced in order to
overcome these difficulties; one doing the tunneling and the other
doing the slow-roll~\cite{LM}.  In this way, the coexistence of two
different mass scales seems more natural. Also, it was argued that in
some models the value of the slow-roll field after bubble nucleation
can be different in each nucleated bubble, and hence the duration of
open inflation would be different in each one.  As a result, for a
given temperature of the CMB, one would obtain a different value of
the density parameter in each universe, and there would always be some
open universes with a density parameter in the interesting 
range~\cite{WV}.

The purpose of this paper is to show that in models of this sort,
with variable $\Omega$, the picture is actually more complicated.
Indeed, instead of an infinite open universe inside of each bubble,
what we find is an infinite number of inflating islands of finite
size inside each bubble.

Quasi-open universes are not entirely new. The simplest two-field
model of open inflation, where the tunneling field $\sigma$ and the
slow-roll field $\phi$ are decoupled, is actually a quasi-open one, as
emphasized by the authors of Ref.~\cite{LM}. Quasi-openness is in
principle not a desirable feature, since to a typical observer, the
universe looks anisotropic~\cite{GM}. In the simple ``decoupled''
model, this ``classical'' anisotropy is large and, combined with the
effect of quantum ``supercurvature'' fluctuations mentioned
above~\cite{misao}, it basically rules out the model~\cite{GM}.

To circumvent this problem, Linde and Mezhlumian introduced a class of
two-field models where the slow-roll field is coupled to the tunneling
field.  As we shall see, these models are also quasi-open. This does
not mean that they are not good cosmological models.  If the co-moving
size of the inflating islands is sufficiently large, then the
resulting classical anisotropy may be unobservable.  Even so, the fact
that these islands are finite leads to a dramatically different
picture of the large scale structure of the universe in open models.

The plan of the paper is the following. In Section II we briefly
review the ``supernatural'' model of inflation, introduced in
Ref.~\cite{LM}.  In this model, an approximate $U(1)$ symmetry is
broken by a tunneling field in a first order phase transition, and
slow-roll inflation inside the nucleated bubbles is driven by the
pseudo-Goldstone boson. In section III, we study the quantum
fluctuations of the pseudo-Goldstone in the bubble background. Section
IV is the core of the paper, where we argue that after tunneling, we
do not obtain an infinite open universe, but an infinite ensemble of
quasi-open universes inside a single bubble. Section V is devoted to
more general models, like the ``coupled'' and ``uncoupled'' two-field
models. In Section VI we briefly describe the observational
implications of our results and in Section VII we summarize our
conclusions. 

\section{Supernatural inflation}

\begin{figure}[t]
\centering
\hspace*{-4mm}
\leavevmode\epsfysize=5.5cm \epsfbox{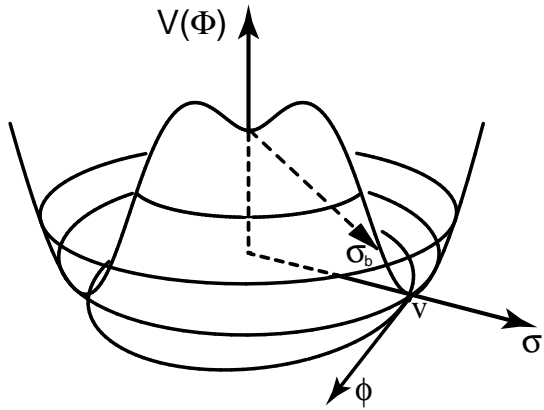}\\[3mm]
\caption[fig2]{\label{fig2} The inflaton potential for the supernatural 
  model. An approximate U(1) symmetry is broken through bubble
  nucleation. We call $\sigma$ the tunneling direction and $\phi$ the
  direction orthogonal to it. The instanton $\sigma_0(\tau)$
  interpolates between false vacuum at $\tau\to \infty$, and
  $\sigma_b$ at $\tau=0$.}
\end{figure}

An attractive scenario for open inflation is the model of a complex
scalar field with a slightly tilted mexican hat potential, see
Fig.~\ref{fig2}, where the radial component of the field does the
tunneling and the pseudo-Goldstone does the slow-roll.  This model was
called ``supernatural'' inflation in Ref.~\cite{LM}, because the
hierarchy between tunneling and slow-roll mass scales is protected by
the approximate global $U(1)$ symmetry.

The action is given by
\begin{equation}
  \label{actionsuper}
  {\cal S} = -\int\,d^4x \sqrt{-g}\,
\Big[\partial_{\mu}\Phi^*\partial^{\mu}\Phi + V(\Phi,\Phi^*)\Big],
\end{equation}
where we use the metric signature $(-,+,+,+)$.
Expanding the field in
the form $\Phi = (\sigma/\sqrt2)\exp(i\phi/v)$, where $v$ is
the expectation value of $\sigma$ in the broken phase, we consider a
potential of the form
$$
V=V_0(\sigma) + V_1(\sigma,\phi)
$$
where $V_0$ is $U(1)$ invariant and $V_1$ is a small perturbation
that breaks this invariance. It is asumed that $V$ has a local minimum
at $\Phi=0$ which makes the symmetric phase metastable.  We shall
consider a `tilt' in the potential of the form $V_1=
\Lambda^4(\sigma)G(\phi)$ where $\Lambda$ is a slowly varying function
of $\sigma$ which vanishes at $\sigma=0$. For definiteness we can take
$G=(1-\cos\phi/v)$.

The idea is that $\sigma$ tunnels from the symmetric phase $\sigma=0$
to the broken phase, landing at a certain value of $\phi$ away from
the minimum of the tilted bottom.  Once in the broken phase, the
potential $V_1$ cannot be neglected, and the field $\phi$ slowly rolls
down to its minimum, driving a second period of inflation inside the
bubble. Another attractive feature of this model is that depending on
the value of $\phi$ on which we land after tunneling, the number of
$e$-foldings of inflation will be different.  Hence it appears that in
principle we can get a different value of the density parameter in
each nucleated bubble. As we shall see, however, this picture is
somewhat oversimplified.

We should point out that the supernatural model is not free from
certain restrictions. Indeed, in order for the pseudo-Goldstone to
realize inflation as in the simple free field ``chaotic'' scenario, we
would need $ v \gtrsim M_p$, where $M_p$ is the Planck mass. On the
other hand, if $V_0$ is a typical quartic potential, the bubble walls
would undergo topological inflation~\cite{TI} for $v \gtrsim M_p$, and
this would spoil the open scenario. Topological inflation occurs when
the thickness of the walls is larger than the Hubble rate at the top
of the potential barrier separating two local minima (degenerate or
not). This is the same condition under which the Coleman-de Luccia
instantons~\cite{CdL} cease to exist, and we have a
Hawking-Moss~\cite{HM} transition instead. Hence the condition $v
\gtrsim M_p$ also represents the regime where the transition is not of
the Coleman-de Luccia type, as it would be necessary for a successful
open universe. As emphasized in Ref.~\cite{LM} in the case of
single-field open inflation, a transition of the Hawking-Moss type
would leave unacceptable anisotropies in the CMB. These constraints
can be made less severe by choosing a suitable form for $V_0$, with
higher curvature at the top of the potential between the two minima,
or perhaps a special form for $G$. In any case, we shall take $v\sim
M_p$ in what follows. 

\section{Quantum fluctuations}

In this section, we compute the amplitude of quantum
fluctuations of the pseudo-Goldstone field in the supernatural
inflation model. For this, we need to review the formalism
for quantizing fields in the background of a bubble.

The field equation for $\Phi$ is
\begin{equation}
\Box \Phi - {\partial V(\Phi)\over \partial\Phi^*} = 0, \label{eom0}
\end{equation}
where $V=V_0+V_1$. In terms of the modulus $\sigma$ and phase
$\phi/v$, we have
\begin{equation}
\Box (\sigma e^{i \phi/v})-\left[{\partial V(\sigma,\phi)\over 
\partial\sigma}
+ i {v \over \sigma} {\partial V(\sigma,\phi) \over \partial\phi}\right] 
e^{i \phi/v}=0. \label{eom}
\end{equation}
It should be noted that classical solutions with $\phi=\tilde \phi=const.$ 
exist only if $\tilde \phi$ is an extremum of $G$, so that
$$\left.\partial V\over \partial \phi \right|_{\phi=\tilde\phi}=0.$$
This is true also for the Euclidean solutions (instantons) describing
the tunneling, so strictly speaking these instantons
can only take us from the false vacuum to the extrema of $V_1$ in the broken 
phase. Even if we relax the condition that $\phi$ should be constant, there 
may not be any instantons which can take us to non extremum values.
Of course, this nonexistence of the corresponding instanton does not mean 
that the field cannot tunnel to a non-extremum value of $\phi$, it
simply means that tunneling away from the extremum will be somewhat 
suppressed. This question will be addressed in Section IV. 

Since the effect we are studying is not due to gravity, we shall start
with the case of a bubble in flat spacetime. Including gravity is
quite straightforward and will be done below. According to the theory
of vacuum decay~\cite{coleman}, the tunneling rate is dominated by the
$O(4)$ symmetric solution of the Euclideanized equations of motion
(\ref{eom}) with appropriate boundary conditions, which is called the
instanton or bounce, and which we shall write as
\begin{equation}
\sigma=\sigma_0(\tau), \quad \phi=\tilde\phi=const, \label{bounce}
\end{equation}
Here, we have introduced the Euclidean radial coordinate $\tau \equiv
({\bf X}^2+T_E^2)^{1/2}$, where $(T_E,{\bf X})$ are cartesian
coordinates in Euclidean space.  As we move from spatial infinity to
the origin, the bounce interpolates between false vacuum
$\sigma_0(\tau\to \infty)=0$ and a certain value of the field in the
basin of the true vacuum, see Fig.~\ref{fig2}, $\sigma_0(0) \equiv
\sigma_b$. In addition, the bounce has to satisfy the boundary
condition $\dot \sigma_0(0) =0$.  The solution describing the bubble
after nucleation is given by the analytic continuation of the
instanton to Minkowski time $T$ through the substitution $T_E= -i T$.
Then, the bubble solution depends only on the Lorentz invariant
`distance' to the origin $({\bf X}^2-T^2)^{1/2}$, where $(T,{\bf X})$
are the usual Minkowski coordinates.

It is useful to change to the new coordinates
\begin{equation}
\tau=({\bf X}^2-T^2)^{1/2},\quad \rho\equiv \tanh^{-1}(T/|{\bf X}|), 
\label{coords}
\end{equation}
in terms of which the line element reads
\begin{equation}
ds^2= d\tau^2 + a_E^2(\tau) d\Omega_{dS^3}; \label{metric1}
\end{equation}
Here $ d\Omega_{dS^3}=-d\rho^2+\cosh^2\rho(d\theta^2+sin^2\theta
d\varphi^2) $ is the line element of a 2+1 dimensional de Sitter space
of unit `radius', and in flat space $a_E(\tau)=\tau$. Including
gravity, $a_E$ has to satisfy the (Euclideanized) Friedmann equation,
as described below.  This 2+1 dimensional de Sitter space can be
thought as the hyperboloid swept by the bubble wall during its time
evolution (suitably rescaled).  In spite of its name, the coordinate
$\rho$ is timelike, whereas $\tau$ is a `radial' spacelike coordinate.

The above coordinates cover only the exterior of the light-cone from
the origin. In order to cover the interior, which is where the open
universe sits, we use the
coordinates
\begin{equation}
t=(T^2-{\bf X}^2)^{1/2},\quad r\equiv \tanh^{-1}(|{\bf X}|/T). 
\label{coords2}
\end{equation}
In terms of these the metric reads
\begin{equation}
ds^2= -dt^2 + a^2(t) d\Omega_{H^3}, \label{metric3}
\end{equation}
where 
$
d\Omega_{H^3}=dr^2+\sinh^2r(d\theta^2+sin^2\theta d\varphi^2)
$
is the metric on the unit 3-dimensional hyperboloid 
and (in flat space) the scale factor is given by $a(t)=t$. 

Notice that (\ref{metric3}) is the metric of an open FRW. When gravity
is included, the scale factor $a(t)$ will no longer be proportional to
the cosmological time $t$ but it will be given by the solution of
the Friedmann equation 
$$(1-\Omega)=(aH)^{-2}.$$ In the general case,
the metrics (\ref{metric1}) and (\ref{metric3}) are
related by the analytic continuation of coordinates and scale factor
in the following way 
\begin{equation}
t=-i\tau,\quad r=\rho + i {\pi\over 2}, \quad a(t)= -i a_E(i t).
\label{analytical}
\end{equation}
These relations can be used to analytically continue solutions 
from the outside to the inside of the light-cone from the origin.

In what follows, we shall assume that tunneling occurs along the 
real direction for $\Phi$, i.e. $\tilde\phi=0$ in (\ref{bounce}), and
we shall consider perturbations around the classical solution of the form
\begin{equation}
\sqrt{2}\Phi=\sigma_0(\tau) + \varphi_1 + i \varphi_2.
\label{perturbations}
\end{equation}
Note that
\begin{equation}\label{phi2}
\phi \simeq v\,{\varphi_2\over\sigma_0}\,.
\end{equation}
Substituting into the action (\ref{actionsuper}) we obtain the second
order action for linearized perturbations
\begin{equation}
\label{pertaction} 
S^{(2)}=S_0[\sigma_0]+
\int\,d^4x \sqrt{-g}[\Box \sigma_0-V_0'(\sigma_0)]\varphi_1 +
S_1[\varphi_1]+S_2[\varphi_2],
\end{equation} 
where
\begin{equation}
S_1 = -{1\over 2}\int\,d^4x \sqrt{-g}
\left(\partial_{\mu}\varphi_1\partial^{\mu}\varphi_1 
+   V_0''(\sigma_0) 
      \varphi_1^2\right), 
\label{s1}
\end{equation}
\begin{equation}
S_2=-{1\over 2}\int\,d^4x\sqrt{-g}      
      \left(\partial_{\mu}\varphi_2\partial^{\mu}\varphi_2 + 
      \left[{V_0'(\sigma_0)\over \sigma_0}+ m^2(\sigma_0)\right] 
      \varphi_2^2\right).
\label{s2}
\end{equation} 
Here
\begin{equation}\label{m2}
m^2(\sigma_0) = {v^2\over\sigma_0^2} \Lambda^4(\sigma_0)\,G''(0) = 
{\Lambda^4(\sigma_0)\over \sigma_0^2} 
\end{equation}
is a small $\tau$-dependent `squared mass' due to the potential
$V_1$. In the last equality we have used $G$ of the form $G=(1-\cos
\phi/v)$. When the field $\sigma$ is in the broken phase, then $m$ is
the mass of the pseudo-Goldstone. Of course, using the unperturbed
equations of motion the linear term in (\ref{pertaction}) drops out.

The action (\ref{s1}) has been studied in some detail in the
past~\cite{JGV,JG} because it is the same as the one for a one-field
model. In particular, it describes the fluctuations of the bubble wall
itself.  Here, we shall concentrate on $S_2[\varphi_2]$, which
describes fluctuations in the direction transverse to tunneling.

In order to study quantum fluctuations, the field $\varphi_2$
is expanded as a sum over modes times the corresponding creation and 
anihilation operators 
\begin{equation}
\varphi_2=\sum \varphi_{plm} a_{plm} + h.c. \label{expansion}
\end{equation}
The equation of motion satisfied by the modes is
$$
\Box \varphi_{plm} - \left[{V_0'(\sigma_0)\over \sigma_0}+ m^2
(\sigma_0)\right] \varphi_{plm} = 0.
$$
Following~\cite{JGV,JG}, we take the ansatz 
\begin{equation}
\label{ansatz}
\varphi_{plm}=a_E^{-1}(\tau) F_p(\tau) {\cal Y}_{plm}(x^i),
\end{equation}
where $x^i=(\rho,\theta,\varphi)$ are coordinates on the 2+1
de Sitter space spanned by the motion of the bubble wall.
Introducing the conformal coordinate $\eta$ defined through the 
relation $a_E(\tau) d\eta = d\tau$,
the equation of motion separates into a Schr{\"o}dinger
equation 
\begin{equation}
-{d^2F_p\over d\eta^2}
 + a_E^2 \left[{V_0'(\sigma_0)\over \sigma_0}+ m^2
(\sigma_0)- {{\cal R}\over 6} \right] F_p = p^2 F_p,
\label{schrodinger}
\end{equation}
where the separation constant $p^2$ plays the role of an energy
eigenvalue, and a Klein-Gordon equation for the modes of a
scalar field of mass $p^2+1$ living in the 2+1 dimensional de Sitter
space
\begin{equation}
\label{bunch}
^{(3)}\Box {\cal Y}_{plm} = (p^2+1) {\cal Y}_{plm}.\label{laplacian}
\end{equation}
Here $^{(3)}\Box$ is the covariant d'Alembertian in this lower
dimensional space, and ${\cal R}$ is the four-dimensional Ricci
scalar, which from the unperturbed Einstein's equations can be written
as ${\cal R}=8\pi G [4V(\sigma_0) + (\sigma_0'/a_E)^2]$.

\begin{figure}[t]
\centering
\hspace*{-4mm}
\leavevmode\epsfysize=5.5cm \epsfbox{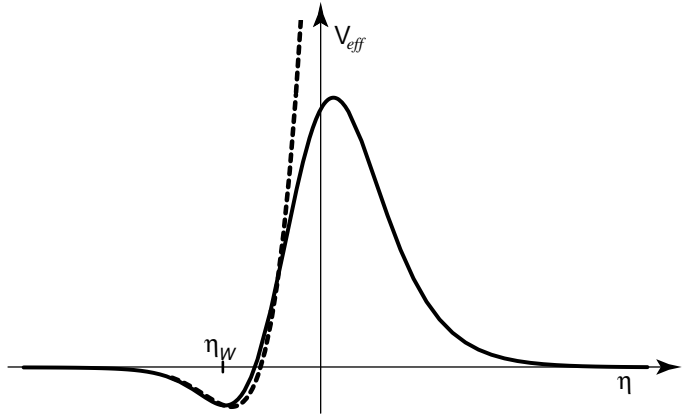}\\[3mm]
\caption[fig3]{\label{fig3} The effective potential in 
  Eq.~(\ref{effpot}). In flat space (dashed line) the potential grows
  without bound at large $\eta$, whereas including gravity it tends to
  zero. At $\eta\to -\infty$, which corresponds to the center of the
  bubble, $V_{\rm eff}$ tends to zero.}
\end{figure}

The modes $\varphi_{plm}$ should be Klein-Gordon normalized on a Cauchy
surface such as $\rho=0$. This amounts to Klein-Gordon normalizing
the lower dimensional modes ${\cal Y}_{plm}$ in the 2+1 dimensional
sense, and then normalizing $F_p$ as in the Schr{\"o}dinger 
problem~\cite{JGV},
\begin{equation}
\int_{-\infty}^\infty F_p F_{p'} d\eta = \delta_{pp'}, \label{norm}
\end{equation} 
where the delta function will be discrete or continuous depending on
whether we are considering discrete or continuous eigenstates of the
Schr{\"o}dinger equation. In flat space, $a_E= R_0 e^{\eta}$, where
$R_0$ is an arbitrary constant which can be conveniently taken to be
of order the radius of the bubble at the time of nucleation.
Therefore, the effective potential in the Schr{\"o}dinger equation
\begin{equation}
V_{\rm eff}(\eta)\equiv  a_E^2 \left[{V_0'(\sigma_0)\over \sigma_0}+ m^2
(\sigma_0)- {{\cal R}\over 6} \right] \label{effpot}
\end{equation}
tends to zero at $\eta\to -\infty$ (center of the bubble) and to
infinity at $\eta\to +\infty$ (false vacuum), see Fig.~\ref{fig3}. In
curved space, it can be shown that~\cite{CdL} $a_E(\eta\to \pm
\infty)\to 0$, so $V_{\rm eff}$ vanishes at both ends.  Therefore, in both
cases, the spectrum will be continuous for $p^2>0$, and there may be a
discrete spectrum for $p^2<0$.

\subsection{Quantum state of a nucleating bubble}

In a time dependent background, the choice of a vacuum state is always
somewhat ambiguous~\cite{qftcs}. Here, this ambiguity corresponds to
the freedom of choosing the `positive frequency' modes ${\cal
  Y}_{plm}$ on the hyperboloid. In principle, the ambiguity can be
resolved dynamically if the initial quantum state before the bubble
nucleates is given.

The quantum state of a nucleating bubble has been extensively studied
both in flat and in curved space~\cite{quantum}. In our model,
$\varphi_2$ is treated as a free field which couples to the bubble via
a $\sigma$-dependent mass term. For this type of models, it has been
shown that if the initial quantum state is de Sitter invariant before
the bubble nucleates, then right after bubble nucleation the field
$\varphi_2$ will be in an $O(3,1)$ symmetric state. This is perhaps
not too surprising: the appearance of the bubble breaks the $O(4,1)$
de Sitter symmetry by selecting a ``nucleation point'' in spacetime,
but otherwise the bubble solution respects an $O(3,1)$ subgroup of
isometries.

What this means is that the positive frequency modes ${\cal Y}_{plm}$
must be taken as the Bunch-Davies modes, which guarantee the desired symmetry.
These are given by
\begin{equation}
{\cal Y}_{plm} = \left[\frac{
\Gamma(l+1-ip)\Gamma(l+1+ip)}{2}\right]^{1/2} 
\frac{P^{-l-1/2}_{ip-1/2}(i \sinh\rho)}{\sqrt{i \cosh\rho}} 
Y_{lm}(\theta,\phi).
\end{equation}
where $P^{\mu}_{\nu}$ are the Legendre functions and $Y_{lm}$ are the
usual spherical harmonics. When analytically 
continued to the inside of the light-cone, through the
relations (\ref{analytical}), they become
\begin{equation}
\label{hyper}
  {\cal Y}_{plm} =
  \left[\frac{\Gamma(l+1-ip)\Gamma(l+1+ip)}{2}\right]^{1/2} 
  \frac{P^{-l-1/2}_{ip-1/2}(\cosh r)}{\sqrt{\sinh r}} Y_{lm}.
\end{equation}
These are proportional to the often used harmonics $Y_{plm}$ 
which are normalized 
on the hyperboloid $H^3$~\cite{super},
$$
{\cal Y}_{plm}={\Gamma(ip) \over \sqrt{2}} \sqrt{{\Gamma(l+1-ip)
\over \Gamma(l+1+ip)}} Y_{plm}.
$$

These analytically continued modes are normalizable on $H^3$ only for
$p^2\geq 0$. Since the mode with $p^2=0$ has wavelength comparable to
the curvature scale, the non-normalizable modes with $p^2<0$ have been
dubbed ``supercurvature'' modes~\cite{super}.  Writing $p=-i \Lambda$,
the supercurvature modes are given by
\begin{equation}
{\cal Y}_{\Lambda, l m}=
\left[{\Gamma(\Lambda+l+1) \Gamma(-\Lambda+l+1)\over2}\right]^{1/2} 
{P_{\Lambda-1/2}^{-l-{1/2}}(\cosh r) \over \sqrt{\sinh r}} 
\,. \label{harmonics}
\end{equation} 
(We have added a comma after the subindex $\Lambda$ to indicate that
it is the value of $ip$ rather than $p$). We repeat, however that if
the corresponding bound state exists in the Schr{\"o}dinger equation
(\ref{schrodinger}), then these modes are perfectly normalizable on
the Cauchy surface, and hence they must be included in the expansion
of the field operator (\ref{expansion}).

\subsection{Degenerate case}

To begin with, let us neglect the mass term $m^2$ (\ref{m2})
which comes from the tilt in the potential, $V_1$ \footnote{Tunneling
  rates in the case when there is an exact internal symmetry have been
  recently investigated in~\cite{KLW}}.  Then, using the equation of
motion satisfied by $\sigma_0$, it is straightforward to show that
\begin{equation}
F_{1}={\cal N}a_E \sigma_0 \label{-1}
\end{equation} 
is a solution of (\ref{schrodinger}) with eigenvalue $p^2=-1$,
or $\Lambda=1$. 
Moreover, this solution is normalizable and it belongs to
the discrete spectrum.
The normalization constant ${\cal N}$ is found from (\ref{norm})
\begin{equation}
{\cal N} = \left(\int_0^{\tau_{max}} a_E \sigma_0^2(\tau) d\tau \right)^{-1/2}.
\label{intN}
\end{equation}
Here, we have changed back to the physical coordinate $\tau$, which
measures the physical distance to the center of the bubble. In flat
space $\tau_{max}$ is actually infinite, but the integral is finite
because $\sigma_0$ vanishes exponentially fast outside the bubble.
Including gravity $\tau_{max}$ becomes a finite value, so the integral
is also finite.  The explicit value of ${\cal N}$ can be calculated
numerically for any given model. If $R_0$ is the size of the bubble
and $\sigma_b$ is the value of the field at the center of the bubble,
see Fig.~\ref{fig1}, at the time of nucleation, then we can estimate
$$
{\cal N} \approx {\sqrt{2}\over R_0 \sigma_b}.
$$
The above estimate is for bubbles such that $R_0$ is small compared
with the Hubble radius, so that $a_E\approx \tau$. Also, we have
substituted $\sigma_0$ by $\sigma_b$ in the integrand, and we have
integrated $\tau$ from zero up to $R_0$. Because $\sigma_0 \leq
\sigma_b$ our estimate is actually a lower bound on ${\cal N}$, and in
some thick wall models the effect can be somewhat higher.

If we asume a large energy difference between
false and true vacua, then we are in the thick wall
regime and $R_0$ is of the order of the thickness of the bubble wall
$$
R_0 \sim M^{-1}.
$$
Here $M$ is the mass of the field in the false vacuum.
If, on the contrary, the vacua are sufficiently degenerate, then we are
in the thin wall regime, and the expression for $R_0$ can be found e.g. in 
Appendix C and Ref.~\cite{thingrav}. 

The normalized mode has an amplitude
\begin{equation}
\varphi_{1,lm}\approx {\sqrt{2}\over R_0} {\sigma_0(\tau)\over\sigma_b}
{\cal Y}_{1,lm}(x_i).\label{fluctuation}
\end{equation}
Physically, what happens is that the field does not simply tunnel to a
sharply defined value of $\phi$, but a distribution of values.
Taking into account that the phase of our complex scalar field $\Phi$
is given by Eq.~(\ref{phi2}), the $x^i$ dependence in
(\ref{fluctuation}) shows that, after nucleation, different points on
the bubble have different values of the angle $\phi$.

When analytically continued to the interior of the light-cone from the
origin, through the relations $(\ref{analytical})$, $\tau$ is replaced
with the cosmological time and $\sigma_0(t)$ quickly follows its
evolution towards its expectation value $\sigma_0 \approx v >
\sigma_b$.  Hence, inside the light-cone, the normalized
supercurvature mode will take the form 
\begin{equation}
\varphi_{1,lm}(t,r,\theta,\varphi)= {\sqrt{2}\over R_0} {v\over\sigma_b}
{\cal Y}_{1,lm}(r,\theta,\phi), \label{rmsdeg}
\end{equation}
which is time independent.  In order for the bubble solution to exist,
the mass of the field in the false vacuum should be $M\gg H_F$, where
$H_F$ is the Hubble rate in the false vacuum. In our model, $H_F$ is
considered to be much greater than the Hubble rate in the true vacuum
$H_T$.\footnote{ We are considering here strongly non-degenerate
  minima. However, in Ref.~\cite{LM} they also consider various depths
  of the central minimum, depending on radiative corrections, and in
  some cases ($g^4=32\pi^2\lambda$) the two minima become degenerate,
  $H_T = H_F$.  This weakens the constraints and makes the model
  viable in certain range of parameters.}  Therefore, the amplitude of
supercurvature perturbations is of order $(v/\sigma_b) M$. This
exceeds by far the amplitude of the usual ``subcurvature''
fluctuations, which is only of order $H_T$. The corresponding effect
in the CMB places severe constraints on the model~\cite{misao}.
We shall come back to this question in Section VI.

Note that for $\Lambda=1$ the amplitude of the homogeneous mode $l=0$
diverges because one of the gamma functions in (\ref{harmonics}) has
vanishing argument. This is related to the fact that, strictly
speaking, the $l=0$ mode should have been quantized as a collective
coordinate instead of as a harmonic oscillator~\cite{KG}. The
divergence simply means that all values of $\phi$ are equally probable
after nucleation.  When we include the effect of $V_1$, the degeneracy
will be broken and the $l=0$ mode will have a finite amplitude.

\subsection{Non-degenerate case}

When the tilt $V_1$ in the potential is included, $p^2=-1$ is no
longer an eigenvalue of the Schrodinger operator (\ref{schrodinger}).
However, it is clear that for small $m^2$ there will still be a
discrete eigenmode whose eigenvalue we can calculate in perturbation
theory. Denoting by $|-1\rangle$ the unperturbed bound state
(\ref{-1}), the perturbation to the eigenvalue will be given by
\begin{equation}
\gamma\equiv p^2+1=\langle-1|a_E^2 m^2|-1\rangle=  
{\cal N}^2 \int a_E^4 \sigma_0^2 m^2(\sigma_0) d\eta.
\label{perturbation}
\end{equation}
Again, $\gamma$ can be computed numerically for any particular model,
but we can estimate it as being of order 
\begin{equation}\label{gamma1}
\gamma =  {\int a_E^3 \sigma_0^2 m^2(\sigma_0) d\tau   
           \over     \int a_E \sigma_0^2(\tau) d\tau}
\sim {1\over2}R_0^2 m_b^2\,,
\end{equation}
where $m_b^2\equiv \Lambda^4(\sigma_b)/\sigma_b^2$, see Eq.~(\ref{m2}).
The above estimate uses the same approximations
as the estimate for ${\cal N}$ after Eq. (\ref{intN}).

The normalized $l=0$ mode will now be given by 
\begin{equation}\label{l=0}
\varphi^{\gamma}_{l=0}\approx {\cal N} \sigma_0 {\cal Y}_{p,00}
\approx  {1\over\pi R_0}{\sigma_0(t) \over \sigma_b}
{1 \over \gamma^{1/2}}{\sinh [(1-\gamma)^{1/2} r]\over 
\sinh r}\ , 
\end{equation}
where, from now on, we shall use the notation $\varphi^{\gamma}_l$ to
denote the modes with $p^2=-1+\gamma$. In this expression, and the
ones that will follow, we give only the unperturbed time dependence,
without including the first order correction in $\gamma$. The field
$\sigma_0(t)$ quickly rolls down from $\sigma_b$ and settles down to
its minimum at $\sigma_0=v$. It is understood that, after that, the
time derivative of the mode will be dominated by the corrections of
order $\gamma$, which slowly drive the Goldstone modes to the minimum
of the tilted potential. The amplitude of the higher-$l$ excitations
will also suffer corrections of order $\gamma$, but since these
amplitudes were already finite, the effect on those is not so dramatic.
To leading order, the amplitudes are the ones calculated in the
previous subsection.

It is interesting to look at the distribution of the field on a
hyperboloid $t=const$. Note that the amplitude of the $l=0$ mode near
the origin $r=0$ is of order
\begin{equation}
\varphi^{\gamma}_{l=0}(r\ll\gamma^{-1})
\approx {1\over \sqrt{2} \pi} {\cal N} 
{\sigma_0 \over \gamma^{1/2}}\approx
{1\over \pi R_0 \gamma^{1/2}} {\sigma_0 \over \sigma_b},
\label{spread2} 
\end{equation}
which is a factor $\gamma^{-1/2}$ larger than the amplitude of the individual
$l>1$ modes found in Eq. (\ref{rmsdeg}). But the amplitude
of the $l=0$ mode decays exponentially for $r\gg \gamma^{-1}$, which means
that at large distances it will become negligible. 
However, the
quantum state that we have chosen is $O(3,1)$ symmetric, which means
that the r.m.s. fluctuation of the field cannot depend on $r$. Therefore,
the loss in amplitude of the $l=0$ mode as we move away from the origin
has to be made up for by the joint contribution of the $l>0$ modes,
smeared over a suitable length scale. This is analogous to what
happens for a massive field in de Sitter space, except that here we
are considering a spacelike manifold $t=const.$ rather than a spacetime.
In the Appendix B we show that for $r\gg \gamma^{-1}$ we have
$$
(\Delta \varphi^{\gamma})^2\equiv
\sum_{l=1}^{l_*} (2l+1) (\varphi_l^{\gamma})^2 \approx
{{\cal N}^2 \sigma_0^2\over 2\pi^2\gamma}\,l_*^\gamma\,e^{-\gamma r} ,
$$
where $l_*\gg 1$ is a certain cut-off.
If we smear over a fixed comoving length $\xi$, as we move away
from the origin we have to include more and more modes in the
sum. Since the $l$-th multipole has wavelength proportional to 
$(\sinh r)/l$, we take $l_*=\sinh r/\xi$. With this, we find
$(\Delta \varphi^{\gamma})^2\approx ({\cal N}^2 \sigma_0^2/2\pi^2 \gamma) 
(2\xi)^{-\gamma}$. Notice that the result is rather
insensitive to the choice of $\xi$. As long as $|\ln \xi|\ll \gamma^{-1}$,
the added contribution of all relevant modes at large $r$ is the same 
as the contribution of the $l=0$ mode near the origin, given by
(\ref{spread2}). We also show in Appendix A that the two-point 
correlations on a $t=const$ surface die off with comoving distance $d$ as 
$e^{-\gamma d/2}$.

Hence, around the time $t_*$ when the field $\sigma_0$ settles down to
its minimum $\sigma_0\sim v$ the supercurvature fluctuations of the
field are of order
\begin{equation}
(\Delta\varphi^{\gamma})
\approx {\sqrt{2}\over \pi}\left({v\over \sigma_b}\right) 
{1\over m_b R_0^2} = {\sqrt{2}\over \pi} 
{\Lambda^2(v)\over \Lambda^2(\sigma_b)} {1\over m R_0^2}  
\sim {M^2\over m}. \label{similar}
\end{equation}
Here, $m=\Lambda^{2}(v)/v$ is the mass of the pseudo-Goldstone in the true
vacuum. In the last step we have asumed that $\Lambda(\sigma)$ is a slowly 
varying
function of $\sigma$, and set $R_0\sim M^{-1}$. Note that the fluctuation in
$\varphi_2$ can easily reach Planckian values. It suffices to take $M\sim
10^{16}$ GeV and $m \sim 10^{13}$ GeV.  With these values, the field is
displaced enough from its minimum that it can drive inflation (recall that we
are asuming $v\approx M_p$).

\section{Quasi-open universe}

Tunneling to a large value of the field is usually understood in an 
``adiabatic'' sense. The idea is that since the motion 
of the phase is dictaded by the explicit symmetry breaking potential $V_1$, 
it will be much slower than the motion of the radial component of the field
dictated by the large U(1) symmetric part of the potential.
Hence, one can estimate the rate for tunneling at any
$\phi$ by solving the Euclidean equations of motion for $\sigma$ 
while $\phi$ is kept as a frozen parameter. This frozen 
parameter is then used as the initial value of the slow-roll field 
on the $t=t_*$ hypersurface. Here, as above,
$t_*$ is the time at which $\sigma_0$ reaches its expectation
value $v$.

\begin{figure}[t]
\centering
\hspace*{-4mm}
\leavevmode\epsfysize=5.5cm \epsfbox{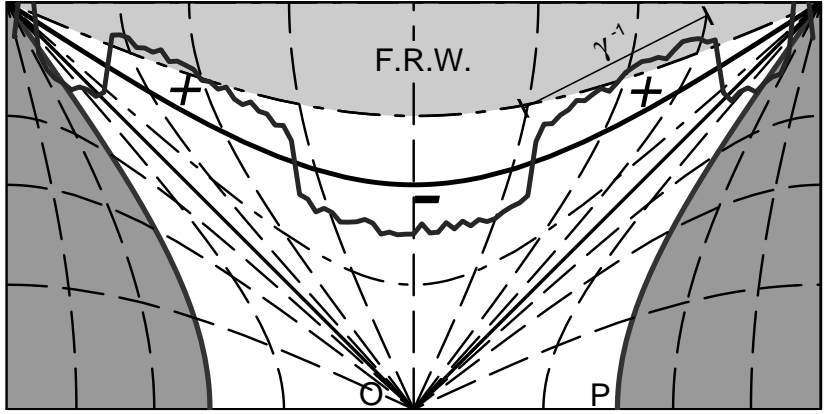}\\[3mm]
\caption[fig4]{\label{fig4} On a $t=const.$ hypersurface inside the 
  bubble, the co-moving coherence length of the slow-roll field is
  $r\sim \gamma^{-1}$.  If the r.m.s. fluctuations of the field are
  large, then regions where the field is large and positive will be
  next to regions where the field is large and negative. These will be
  separated by regions where the field is small and there is no second
  stage of inflation.}
\end{figure}

However, it is clear that when the spread of the pseudo-Goldstone on
the $t=t_*$ hyperboloid is comparable to its range, $(\Delta
\varphi^{\gamma})\gtrsim \pi v$, the picture that each bubble
nucleates with a different value of $\varphi_2$ is not adequate.
Instead, all of the `vacuum' manifold is pretty much sampled {\em
  inside a single bubble}. In this case, inflation does not take place
coherently on the $t=const.$ hyperboloids. As mentioned in the
previous Section, the comoving correlation length for quantum
fluctuations (in units of the curvature scale) is $d\sim\gamma^{-1}$.
Thus, patches of comoving size $d\sim \gamma^{-1}$ where $\varphi_2$
is large and positive would be right next to patches where $\varphi_2$
is large and negative. These patches would be separated by regions
where $\varphi_2$ is small and the universe does not inflate.  Patches
with positive and negative values can also be separated by ``walls''
where $\varphi_2$ is close to $\pi v$. In the case $v\approx M_p$
these domain walls of the pseudo-Goldstone potential would be
topologically inflating~\cite{TI}. Hence, instead of a smooth
inflating hyperboloid, we have a patchy mosaic of inflating regions,
as depicted in Fig.~\ref{fig4}. In principle, each patch can give rise
to a successful cosmology, but this cosmology will not be an open
universe in the traditional sense. At best it will be a `quasi-open'
one, i.e.  one which locally resembles an open FRW.

When $(\Delta \varphi^{\gamma})\ll \pi v$ the spread in the
distribution of the pseudo-Goldstone in a $t=t_*$ section is small.
What this means is that if the nucleated bubble is described by the
$O(3,1)$ symmetric quantum state, then most of the surface $t=t_*$ has
a non-inflating value of the field.  However, in an infinite
hypersurface, there will be a certain density of occasional large
fluctuations which will lead to inflating islands of comoving size
$d\sim\gamma^{-1}$.  Each one of these islands will be a quasi-open
universe.

This is in clear opposition with the conventional adiabatic picture
described in the first paragraph. It is interesting to pursue the
adiabatic picture for a moment, in order to see in which sense it is
adequate or not. To keep the discussion simple, let us consider
tunneling to a range of values of $\phi$ for which the linearized
expressions (\ref{pertaction}) are still valid, but sufficiently large
that it can be distinguished from tunneling to the bottom $\phi=0$.
This will be the case if $\varphi_2$, see Eq. (\ref{phi2}), is
in the range
\begin{equation}
\label{range}
(\Delta\varphi_2) \ll  \varphi_2 \ll  v,
\end{equation}
where $(\Delta\varphi_2)^2=({\cal N}^2\sigma_0^2/2\pi^2\gamma)$ was 
computed in the previous section.

Since $\varphi_1$ and $\varphi_2$ decouple, we may take an {\em
  approximate} Euclidean solution of the form
$\Phi=(1+i\phi/v)\sigma_0(\tau)/\sqrt{2}$, where $\phi$ is taken as
constant in the adiabatic approximation.  Substituting this
configuration in the Euclideanized version of (\ref{pertaction}) we
find, after straightforward algebra,
$$
S_E[\phi]=
S_E[\sigma_0]+{1\over 2}\int d^4x \sqrt{-g}\,m^2\varphi_2^2 = 
S_E[\sigma_0]+{\gamma\pi^2 \over {\cal N}^2 v^2} \phi^2. 
$$
If the decay rate is proportional to $e^{-S_E}$, the relative
probability of having a bubble with a certain value of $\varphi_2$ at
nucleation will be
\begin{equation}
\label{probability} 
{\cal P}(\varphi_2)\sim \exp\left[{-\varphi_2^2\over 
2 (\Delta\varphi_2)^2}\right],
\end{equation}
where we have used (\ref{phi2}) and, perhaps not too surprisingly, 
found $(\Delta\varphi_2)^2=({\cal N}^2 \sigma_0^2/2\pi^2\gamma)$, the
same expression obtained in the previous section from considerations of
quantum fluctuations in the $O(3,1)$ invariant state.
Thus, two approaches which in principle are aimed at answering different
questions end up giving the same answer. Here, we were asking how likely
is it for a bubble to nucleate at a large value of $\phi$, whereas
in the previous Section
we were computing the amplitude of fluctuations inside a given bubble. 
Undoubtedly, both questions are related, since at least locally we cannot 
distinguish a large value of the field induced by the nucleation of the bubble
from a fluctuation of the field inside the bubble.

However, even though the adiabatic approximation may give the right
answer for the probability of tunneling to a large value of $\phi$, it
suggests the wrong picture for bubble nucleation.  Since $\phi=const$
is used in the above estimate, one might imagine that an infinite open
universe with homogeneous $\varphi_2$ in the range (\ref{range}) can
be created. We shall argue that the probability for this to happen is
actually zero.  First of all, the formula $P\sim \exp[-S_E]$ is only
justified when the Euclidean action is evaluated on a solution of the
equations of motion. But $\phi=const$ is only a solution in the case
when $V_1$ is neglected. We can try and correct this configuration so
that it will be a solution, while still keeping $O(4)$ symmetry. In
the linear regime, what this means is that we want a solution of Eq.
(\ref{schrodinger}) with $p^2=-1$, so that Eq. (\ref{laplacian}) is
satisfied by ${\cal Y}_{plm}(x^i)=const.$ and we have a homogeneous
solution inside the bubble.  However, for $m^2>0$ the lowest
eigenvalue is $p^2=-1+\gamma>-1$, which means that the solution with
$p^2=-1$ is not normalizable.  As a result, the corresponding
Euclidean action is badly divergent. If the action is regularized with
a cut-off and if we take the $p^2=-1$ solution to be well behaved at
the center of the bubble $\tau=0$, then it is easy to show that the
action starts growing exponentially, $S_E\sim \exp[2M(\tau_c-R_0)]$,
as the cut-off $\tau_c$ in the radial direction $\tau$ becomes larger
than the size of the bubble. Here $M$ is the mass of $\Phi$ outside
the bubble.  Hence it is not justified to say that the estimate
(\ref{probability}) gives the nucleation rate for a homogeneous
bubble. Rather, using a homogeneous solution we would get a divergent
action and hence a vanishing probablility.

\subsection{Creation of a quasi-open universe}

To compare, we can now ask what is the amplitude for tunneling
from false vacuum to a spherically symmetric but inhomogeneous
configuration with a large value of $\varphi_2$ inside the bubble.
This is what we call an ``inflating island'' or quasi-open universe.
Again, this amplitude will depend on the
action of a semiclassical Euclidean trajectory.
In order to make the metric (\ref{metric3}) into one of
Euclidean signature, we must consider the analytic continuation
of the coordinates $t=- i\tau$ and $r=ir_E$. With this we have
\begin{equation}
ds^2=d\tau^2+a_E^2(\tau)d\Omega_{S^3}, \label{metriceucl}
\end{equation} 
where $d\Omega_{S^3}=dr_E^2+sin^2r_E(d\theta^2+sin^2\theta d\phi^2)$ 
is the metric on the three-sphere, and
the range of $r_E$ is from
$0$ to $\pi$. 

The semiclassical trajectory we shall consider is simply the analytic
continuation of the $l=0$ supercurvature mode \footnote{The
  semiclassical trajectory we consider does not have vanishing
  ``temporal'' derivative at the bounce point $r_E=\pi/2$ where we
  match the Euclidean solution to the Lorentzian one.  Hence it is a
  complex trajectory which has a small imaginary part (of order
  $\gamma$) in the classically forbidden region.  The imaginary part
  decreases exponentially fast in the Lorentzian section on a
  timescale of order $R_0^{-1}$} (\ref{l=0})
\begin{equation}
\varphi_2 = A\  f(\tau) g(r_E),
\label{inhomogeneous}
\end{equation}
where $f(\tau)=\sigma_0(\tau)+O(\gamma)$ and 
$$
g(r_E)={\sin[(1-\gamma)^{1/2}r_E]\over\sin r_E.}
$$
Note that this solution is not regular at one of the poles of the
three-sphere, $r_E=\pi$. This is of course expected, since $g$ is an
eigenfunction of the Laplacian with eigenvalue $\gamma$, which is not
in the spectrum. Hence, (\ref{inhomogeneous}) is not a regular
``bounce'' around which we should expand in order to find the false
vacuum persistence amplitude~\cite{coleman}.\footnote{This is a
  blessing, because an inhomogeneous instanton would cause the well
  known problem that the decay rate should be multiplied by the
  infinite volume of the Lorentz group.} This is not a problem for us
here, because we are not estimating the total decay rate of the false
vacuum but a particular transition amplitude.  For this, we only need
``half'' of the Euclidean solution, which interpolates between false
vacuum at $\tau \to \infty$, $r_E\approx 0$ and an inhomogeneous
configuration at $r_E= \pi/2$. At $r_E=\pi/2$ this solution is matched
to a Lorentzian solution at $\rho=0$ (the spacelike surface where the
bubble nucleation takes place) simply by the analytic continuation
$t=-i\tau$ and $\rho=i(r_E-\pi/2)$ [see Eq. (\ref{analytical})]. The
solution is then propagated to the interior of the bubble (the
``open'' universe) through Eq. (\ref{analytical}).

The transition amplitude $\Psi$ is WKB suppressed only in the 
Euclidean regime; the Lorentzian evolution contributing 
an oscillatory phase. Therefore
$$
|\Psi|\sim e^{-S_E},
$$
where 
$$
S_E= 2\pi \int_0^{\infty}d\tau\ a_E^3(\tau)\int_{0}^{\pi/2}
dr_E\ \sin^2r_E\left[\dot\varphi_2^2+{1\over a_E^2}
\left({\partial \varphi_2\over \partial r_E}\right)^2+
\left({V_0'\over\sigma_0}+m^2\right) \varphi_2^2 \right].
$$
It is easy to show that since $\varphi_2$ is a solution of
the equations of motion, after integration by parts
the only contribution to the 
integral comes from the boundary term at $r_E=\pi/2$. A straightforward
calculation gives
$$
S_E=A^2\,{\gamma\pi^2\over 2}\int\  a_E^2\sigma_0^2 d\eta\approx
    A^2\,{\gamma\pi^2\over 2 {\cal N}^2},
$$
where $A$ is de constant introduced in (\ref{inhomogeneous}).
After analytic continuation to the interior of the bubble,
we have $\varphi_2(r\ll \gamma^{-1})\approx A\sigma_0$.
Hence, the relative probability for nucleating through the
inhomogeneous
trajectory (\ref{inhomogeneous}) is again given by
\begin{equation}
{\cal P}=|\Psi|^2\sim \exp\left[{-\varphi_2^2\over 2 (\Delta\varphi_2)^2}
\right]. \label{creation}
\end{equation}
Although this is in perfect agreement with (\ref{probability}),
it is now clear that it doesn't mean the whole open universe will
have the value $\varphi_2$, but only a patch of comoving 
size $\sim \gamma^{-1}$ will have this value.

Thus, tunneling to a value of the field which is far from the one
indicated by the $O(4)$ symmetric instanton is perfectly possible,
with a somewhat suppressed probability. However, the
resulting universe is not an infinite open universe but just a 
quasi-open one.

\subsection{Many universes in one bubble}

The arguments used in the previos subsection leading to
Eq.~(\ref{creation}) are somewhat heuristic. In particular, we have
not attempted to justify why the semiclassical trajectories of the
form (\ref{inhomogeneous}) should be the only relevant ones. Note,
however, that the probability distribution (\ref{creation}) for
nucleating at a high value of the field $\varphi_2$ near $r=0$, is the
same as the Gaussian distribution for the amplitude of the $l=0,
p^2=-1+\gamma$ mode in the $O(3,1)$ invariant state. In fact, the
possibility of nucleating at different values of the field is already
accounted for by this quantum state and need not be considered
separately. The analysis of Refs.~\cite{quantum}, whose result we
described in Section III.A, takes into account all paths in
$\phi$-field space, not just the semiclassical one used above. That
analysis should be regarded as a more rigorous derivation of the
result (\ref{creation}).

Because of the invariance of the quantum state, it is clear that there
is nothing particular about the point $r=0$, and an inflating `blob' is
equally likely to develop around any point on the $t=t_*$ hyperboloid.
Therefore we are led to the picture where in each comoving volume of
size comparable to the correlation lenght $\gamma^{-1}$, the
probability distribution for $\varphi_2$ is given by (\ref{creation}).
Clearly, since the volume of the hyperboloid is infinite, inflating
islands with all possible values of the field at their center will be
realized inside of a single bubble. We may happen to live in one of
those patches of comoving size $d \lesssim \gamma^{-1}$, where the
universe {\em appears} to be open.

Also, in the supernatural model, it is possible to modify the shape of
the potential near the false vacuum so that there is a misalignment
between the prefered direction for tunneling~\cite{LM} and the
direction of the minimum of the pseudo-Goldstone potential in the
broken phase. In this case, we expect that the $t=t_*$ surfaces inside
the bubble will have a mean value of $\phi=\phi_c \neq 0$, determined
by the most probable escape path (i.e., the instanton, which in this
case will not land on $\phi=0$).  This value of $\phi_c$ will
determine the number of $e$-foldings of inflation and hence the mean
value of the density parameter $\Omega_0$ on the hyperboloid. Let us
call this value $\Omega_c$. If the tunneling path is not too narrow,
there will still be a supercurvature mode which will cause
fluctuations in the density parameter on co-moving scales of order
$\gamma^{-1}$, which are of course much larger than the Hubble radius.
The picture is then that we have an ensemble of large patches with
different values of the density parameter. This is an interesting
situation which deserves further study. However, since this model
involves more parameters, for the remainder of this paper we shall
concentrate on the simplest case discussed above, where the preferred
tunneling direction and the minimum of the pseudo-Goldstone potential
are aligned.

\section{More general models}

In this section we shall consider the class of two-field models with
a potential of the form 
\begin{equation}
V(\sigma, \phi)= V_0(\sigma)+{1\over 2}m^2\phi^2 + 
{1\over 2} g \sigma^2\phi^2.
\label{general}
\end{equation}
Models of this type were introduced in Ref.~\cite{LM}.  Here $V_0$ is
a non-degenerate double well potential, with a false vacuum at
$\sigma=0$ and a true vacuum at $\sigma=v$.  When $\sigma$ is in the
false vacuum, $V_0$ dominates the energy density and we have an
initial de Sitter phase with expansion rate given by $H_F^2\approx
(8\pi G/3) V_0(0)$.  Once a bubble of true vacuum $\sigma=v$ forms,
the energy density of the slow-roll field $\phi$ may drive a second
period of inflation.

As pointed out in Refs.~\cite{LM}, the simplest two-field model of
open inflation, given by (\ref{general}) with $g=0$ and $m\neq 0$,
is actually a quasi-open one. Since there is no coupling between
the two fields except for the gravitational one, we shall call this
the ``decoupled'' two-field model. In this model, inflation starts
chaotically at large values of $V(\sigma,\phi) \lesssim M_p^4$. In
some regions of the universe, the field $\sigma$ will be trapped in
the false vacuum, while $\phi$ rolls down from large values. If a
bubble nucleates at a point where $\phi\sim M_p$, the value of the
slow-roll field will be large enough to drive a short second period of
inflation inside the bubble. One problem with this model, is that the
slow-roll field moves also outside the bubble, so the synchronization
of the $\phi=const.$ and $\sigma=const.$ surfaces inside the bubble is
not perfect, as pointed out in Ref.~\cite{LM}.

In Ref.~\cite{GM}, the classical evolution of the slow-roll field from
the outside to the inside of the bubble was studied, and it was found
that on the hypersurface $t\sim H_F^{-1}$ the high value $\phi\sim
M_p$ decays exponentially with the distance to the origin as
\begin{equation}
\phi\propto \exp[-\gamma_c r/2] \label{classical}
\end{equation}
where $\gamma_c \approx {2 m_F^2/ 3H_F^2}$. Hence, the size of the
inflating region in this model is finite. We use the subindex $c$ to
stress that this result follows from purely classical evolution.  The
larger $H_F$, the larger will be the inflating region.  The reason is
that the cosmological friction term in the equation of motion for
$\phi$ is proportional to $H_F$, so the larger is $H_F$, the slower the
field $\phi$ will roll down the potential outside the bubble, and the
better is the synchronization between $\phi=const.$ and
$\sigma=const.$ surfaces. However, $H_F$ cannot be taken to be too
large because otherwise the quantum fluctuations generated outside the
bubble produce too large an amplitude for the supercurvature mode
inside the bubble~\cite{misao}.  The combination of these two effects
severely constrains this model~\cite{GM}.

In order to construct a truly open model, Linde and Mezhlumian
suggested taking $m=0$ and $g\neq 0$. We shall call this the
``coupled'' two-field model. In this way, the mass of the slow-roll
field vanishes in the false vacuum, and it would appear that the
problem of classical evolution outside the bubble is circumvented.
However, this is not exactly so, and the whole class of models
(\ref{general}) leads to quasi-open universes. The basic reason is
that, as we shall see below, the (linear) equation of motion 
for $\phi$ in the presence of the bubble, does not admit
$O(3,1)$ invariant solutions which are regular at the origin, except
for the trivial one, $\phi=0$. Thus, we are back to a situation
analogous to the supernatural inflation model.

\begin{figure}[t]
\centering
\hspace*{-4mm}
\leavevmode\epsfysize=5.5cm \epsfbox{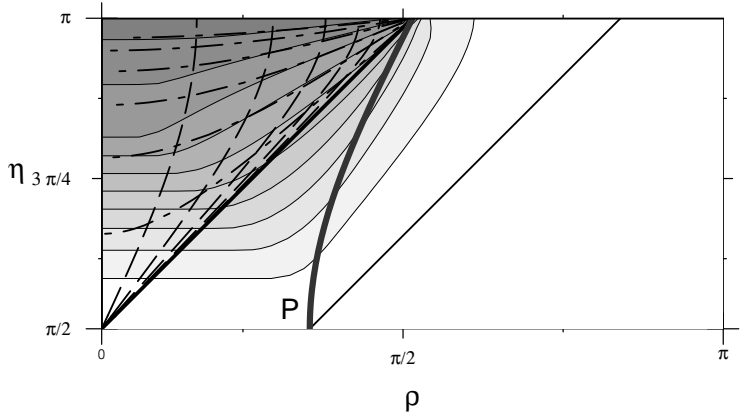}\\[3mm]
\caption[fig5]{\label{fig5} Conformal diagram of de Sitter space with 
  a bubble expanding in it. Here $\eta$ and $\rho$ are the usual
  conformal coordinates in the {\em closed} chart.  The bubble wall is
  represented by the curved thick line starting at the point P. The
  figure shows the result of a numerical evolution of the slow-roll
  field $\phi$ in the coupled model (\ref{general}) with $m_F=0$ and
  $m_T=1.5 H_T$.  The self-gravity of the bubble has been ignored and
  we have taken $H_T=H_F$. The initial conditions for $\phi$ at
  $\eta=\pi/2$ are $\phi=const$ and $\dot\phi=0$. In spite of the fact
  that the field is massless outside the bubble, it starts evolving
  everywhere inside the light-cone from the point P. As a result, the
  surfaces of constant $\phi$, separating regions with different
  shadings, are not well synchronized with the $t=const.$ surfaces
  inside the bubble (see Fig.~\ref{fig1}). In the plot, the field
  decays by one tenth of its initial value between consecutive
  $\phi=const.$ lines.}
\end{figure}

Even if the mass of the field in the false vacuum vanishes, one must
not expect that $\phi$ will not evolve at all outside the bubble.
Fig.~\ref{fig5} shows the result of a numerical evolution of the field
$\phi$ in the coupled model. The figure represents a conformal diagram
of a bubble expanding in de Sitter space (for simplicity, the
gravitational field of the bubble has been neglected).  The bubble
wall is indicated by the thick timelike hyperbola.  As initial
conditions, we have taken $\phi=const.$ and $\dot \phi=0$. Surfaces of
constant $\phi$ are indicated by different shadings. Even though the
field is massless outside the bubble, we find that it does not stay
exactly constant there.  Due to the finite size of the bubble at the
time of nucleation, the field $\phi$ feels the presence of the bubble
everywhere inside the light-cone from the ``point'' P. As a result,
inside the bubble, the hypersurfaces $\phi=const.$ are not perfectly
synchronized with the $\sigma=const.$ hypersurfaces. Thus, we are back
to a situation where the inflating region inside the bubble has a
finite size, as in the decoupled model (note that the $\phi=const.$
lines cross the bubble wall trajectory).

Note that this effect is due to the finite size of the bubble.  We
shall see below that the effect is of order $(H_F R_0)^4$.  In
Fig.~\ref{fig5}, the parameters have been chosen so that the effect is
very dramatic and the size of the inflating islands is comparable to
the curvature scale, but one can choose parameters so that the
inflating islands are as large as desired. However, except in the case
where gravity is neglected so that $H_F=0$, their size is always
finite and the large value of the field at the time of nucleation ends
up decaying at large distances $r$ from the origin.

Hence, just as in the case of the supernatural model, the infinite
$t=const.$ surfaces would be almost empty at large distances, if it
wasn't for the occasional quantum fluctuations which may ignite
inflating islands here and there.

\subsection{Quantum fluctuations}

As in the case of the supernatural model,
we expand the field operator $\phi$ in terms of
creation and anihilation operators,
\begin{equation}
\phi=\sum \varphi_{plm} a_{plm} + h.c. \label{expansion2}
\end{equation}
In the present case, the equation of motion for the modes is given by
$$
\Box \varphi_{plm} - [m^2 + g \sigma_0^2] \varphi_{plm} = 0.
$$ 
Here $\sigma_0$ is the ``background'' bubble solution.
Using the ansatz (\ref{ansatz}), 
$$
\varphi_{plm}=a_E^{-1}(\tau) F_p(\tau) {\cal Y}_{plm}(x^i),
$$
we have the following
Schr{\"o}dinger equation for $F_p$,
\begin{equation}
-{d^2F_p\over d\eta^2}
 + a_E^2 \left[m^2+ g \sigma_0^2- {{\cal R}\over 6} \right] F_p = p^2 F_p.
\label{schrodinger2}
\end{equation}
This equation determines the spectrum of allowed eigenvalues $p^2$, 
which correspond to normalizable eigenfunctions $F_p$. All of these
eigenvalues have to be included in the expansion of the field operator  
(\ref{expansion2}). As before, the harmonics ${\cal Y}_{plm}$ must satisfy
equation (\ref{laplacian}).

In the case of supernatural inflation, we saw that there was a
discrete eigenstate with $p^2<0$, which actually dominated the r.m.s.
fluctuations of the field on a $t=const.$ hypersurface.  We shall see
that a similar situation happens in this case.

Since the Hubble rate inside the bubble $H_T$ is smaller than the
Hubble rate outside, $H_F$, we need
\begin{equation}
m^2+g v^2 \ll H_T^2 < H_F^2 \label{condition}
\end{equation}
in order to have slow-roll inflation inside the bubble.  This
condition suggests taking a perturbative approach. To lowest order, we
can neglect the mass term for $\phi$ and the gravitational
backreaction of the bubble, so that $\phi$ is just a massless field in
de Sitter space. In this case, Eq. (\ref{schrodinger2}) has the well
known supercurvature mode with $p^2=-1$, which corresponds 
to~\cite{super}
\begin{equation}
F_{-1}= {H_F\over \sqrt{2}} a_E(\eta)
\label{usual}
\end{equation}
The field configurations corresponding to this $p$ are
$\phi=const.\times {\cal Y}_{1,lm}$. As mentioned before, for $l=0$
the harmonic ${\cal Y}_{1,00}$ is not Klein-Gordon normalizable in the
2+1 dimensional sense. This is because, for a massless field in de
Sitter space, the zero mode corresponding to the translations $\phi\to
\phi+ const$ has to be treated as a collective coordinate and not as
an oscillator. When the mass of $\phi$ is included, the mode becomes
normalizable. Indeed, the bound state will shift to a perturbed
eigenvalue $p^2=-1+\gamma$ which, as before, can be calculated
perturbatively:
\begin{equation}
\gamma\equiv p^2+1=  {H_F^2\over 2} \int a_E^4
 \left[m^2+ g \sigma_0^2- {{\cal R}\over 6} + 2 H_F^2 \right] d\eta.
\label{perturbation2}
\end{equation}
The normalized modes corresponding to the discrete eigenvalue take the
form $\varphi^{\gamma}_{lm}\approx (H_F/\sqrt{2}) {\cal
  Y}_{plm}(x^i)$, and in particular, for the $l=0$ mode,
\begin{equation}
\varphi^{\gamma}_{l=0}\approx {H_F\over 2\pi} {1\over \gamma^{1/2}}
{\sinh [(1-\gamma)^{1/2} r]\over \sinh r}[1+O(\gamma)].\label{zero2}
\end{equation}
The uncertainty of order $\gamma$ comes from the fact that we have not
evaluated the correction to the ``wave function'' $F_{-1}$ which gives
the temporal dependence of the field. This can be done in principle,
but it is not really necessary for our purposes.  It is clear that the
mass term will cause the field to have the temporal dependence
corresponding to slow-roll inside the bubble.

Near the origin $r=0$ the r.m.s. fluctuation of the 
field $\Delta \phi$ will be dominated by the mode (\ref{zero2}), and
for $t=t_*\sim H_F^{-1}$ it will be given by
\begin{equation}
\Delta \phi \approx {H_F\over 2\pi} {1\over \gamma^{1/2}}.
\label{rms}
\end{equation}
As mentioned in Section III, because of the $O(3,1)$ invariance of the
quantum state, this will also be the r.m.s. fluctuation of the field
at any point on the $t=t_*$ hypersurface (see Appendix B).

In the case of thin walls, the value of $\gamma$ can be calculated
explicitly (see Appendix C),
\begin{equation}
\gamma = {2\over 3}{m_F^2\over H_F^2}+
             {1\over 8} H_F^2 R_0^4 (m_T^2 -m_F^2).
\label{gamma}
\end{equation}
Here $m_T^2= m^2 + g v^2$ is the mass of the slow-roll field in the true
vacuum, $m_F^2=m^2$ is the mass in the false vacuum, $H_F$ is the
hubble rate in the false vacuum, and $R_0$ is the intrinsic radius of the
bubble at the time of nucleation.

The origin of the different terms in (\ref{gamma}) is easy to
understand.  The first one is independent of the existence of the
bubble, and comes from the fact that the slow-roll field has a mass
$m_F^2$ in the false vacuum (the $H_F^2$ in the denominator can be
understood from simple dimensional considerations). The second term is
due to the perturbations of the effective potential in the
Schr{\"o}dinger equation (\ref{schrodinger2}) caused by the bubble
solution.  In the bubble, the scale factor $a_E$ is of order $R_0$, so
the factor $a_E^4$ in the integrand of (\ref{perturbation2}) will
yield the factor $R_0^4$ in front of the second term of (\ref{gamma}).

\subsection{Inflating islands}

Even though the decoupled model ($g=0$) is not a very good candidate
to an open cosmological model~\cite{GM}, it is instructive to consider
it as a first step. In this model, inflation inside the bubble can be
initiated because of the large
``classical'' value of $\phi$ at the time of nucleation, but at very
large distances from the origin $r \gg \gamma^{-1}$, the classical
field dies off, and only the quantum fluctuations remain.

Let us consider the amplitude of quantum fluctuations. From
(\ref{gamma}) we obtain ($m_F= m_T$) 
$$
\gamma = {2 m_F^2\over 3H_F^2}.
$$
Accordingly, from (\ref{rms}), the r.m.s. quantum 
fluctuations of the field on the $t=t_*$
surfaces will be of order 
\begin{equation}
(\Delta \phi) \approx \sqrt{3\over 2}{1\over 2\pi} {H_F^2\over m_F},
\label{great}
\end{equation}
and the comoving correlation length will be of order 
$\gamma^{-1}=\gamma_c^{-1}$, the same as we found in the classical case
(\ref{classical}) from a completely different approach. 

In the decoupled model $m_F$ is the same as $m_T$, and $H_F$ is not
too much larger than $H_T$~\cite{LM,misao,GM}.  Since $H_T\sim m_T$, we
have $(\Delta \phi) \ll M_p$ and quantum fluctuations on that surface
would typically not reach ``inflating'' values of order $M_p$.  Still,
as in the supernatural model, an occasional large quantum fluctuation
can initiate inflation on a patch of size $\gamma^{-1}$.

Let us reflect upon the meaning of Eq. (\ref{great}). This r.m.s.
amplitude is actually the same as the Bunch-Davies one for
fluctuations of the slow-roll field outside the bubble~\cite{BD}.
Hence, the correct interpretation of the result (\ref{great}) seems to
be the following. Inflation inside and near the bubble wall may start
because the field is large at the point where the bubble nucleates.
However, even after bubble nucleation, the field will continue its
random walk outside the bubble, and it may occasionally become large.
If the bubble wall hits a patch where the field is large, then this
will generate a local inflating patch inside the bubble, and we might
inhabit one of those inflating patches. However, this model is not a
very good open model, and it would only agree with observations if
$\Omega_0$ were very close to one. 

Let us now consider the ``coupled'' model, where the fields are coupled
but $m_F=m=0$. In this case 
$$
\gamma \approx {1\over 8} m_T^2H_F^2 R_0^4.
$$
Clearly, by choosing parameters such that the size of the bubble is
much smaller than the Hubble rate outside the bubble, or such that the
mass of the field in the true vacuum is sufficiently small, $\gamma$ can
be made as small as desired. Hence the size of inflating regions can
be made as large as desired. In this case, the field is massless
outside the bubble and quantum fluctuations of $\phi$ pile up to
arbitrarily large values far from the bubble.  However, from Eq.
(\ref{rms}), we find a finite answer for the fluctuations inside the
bubble
\begin{equation}
\label{finite}
(\Delta \phi) \approx {\sqrt{2}\over \pi}{1\over m_T R_0^2}.
\end{equation}
The first interesting thing to note about this result is that it does
not depend explicitly on the Hubble rate inside or outside the bubble
(the only dependence is through $R_0$).  The second observation is
that it is very similar to the expression (\ref{similar}) we had for
the supernatural case, so a connection between the physics of both
models can be anticipated.  The finiteness of (\ref{finite}) is not
surprising, since the slow-roll field is coupled to the bubble, and
piling of modes in the vicinity of the bubble is suppressed by the
mass term. Also, nucleation of bubbles at high values of $\phi$ is
suppressed because the degeneracy between true and false vacuum is
lower. As we discussed in Section IV, the quantum state already
encodes the information that tunneling to a large value of the field
is suppressed.

A difference with the supernatural inflation case is that now the
amplitude of the supercurvature modes with $\l \neq 0$ is of
order $H_F$ rather than $1/R_0$, hence the constraints
on this model from microwave background anisotropies will be easier 
to accomodate.

\section{Observational consequences}

Our results from the previos sections have important consequences for
two-field models of open inflation.  First of all, our models are
quasi-open, rather than open, which leads to classical
anisotropies~\cite{GM}. Second, we saw that in the supernatural
model, the amplitude of supercurvature excitations is quite large. In
this section, we shall give order of magnitude estimates for the
expected CMB anisotropies from these effects.  A detailed
investigation of the power spectrum will be presented
elsewhere~\cite{GBGM}.

\subsection{Classical anisotropies}

Quasi-open universes are finite, and hence they look anisotropic
to a typical observer. This effect was studied in Ref.~\cite{GM} 
for the uncoupled model, and was called a ``classical
anisotropy''. The name was given because the finiteness of
inflating islands was due to the classical motion of
the slow-roll field outside of the bubble. 
Clearly, the same
effect arises in all quasi-open universes we have considered. 
In some cases the appearance of the inflating island is better
described as a semiclassical effect, but the resulting inflating islands
are just as classical here as they were in Ref.~\cite{GM}. Hence, we
shall use the same name for this type of anisotropies.

To proceed, it will be important to distinguish between two different
cases. The first case arises when the r.m.s. fluctuation of the
slow-roll field $\phi$ on the spacelike surfaces $t=t_*\sim H_F^{-1}$
is small compared with $M_p$. In this case, ``high peaks'' where the
field is comparable to $M_p$ and which will lead to inflating regions
of co-moving size $r\sim\gamma^{-1}\gg 1$ will be very ``rare'' on
that hypersurface.  Here, $\gamma$ is the correction to the
supercurvature eigenvalue, calculated in Section III.C for the
supernatural model and in Section V.A for the more general models.
High peaks of a homogeneous Gaussian random field tend to be
spherical, and so our inflating islands will have approximate
spherical symmetry. In the opposite case, when the r.m.s. fluctuation
of the slow-roll field is comparable to $M_p$, we will have a patchy
mosaic of ``overlapping'' inflating islands, as described in the
second paragraph of Section IV. It is easy to check that in the second
case the ``classical'' effect is small compared with the effect of
quantum fluctuations which we shall consider in the next subsection,
so here we shall only consider the first case.

Let us begin with the supernatural model.  The quantum state we are
considering leads to a Gaussian distribution for the random field
$\phi$ that is $O(3,1)$ symmetric. Hence, to compute the probabilities
for the field distribution around any point it suffices to study them
around the origin $r=0$.  Here we shall only be concerned with
fluctuations due to the supercurvature modes, which have a long range.
The effect of subcurvature modes can be incorporated in the usual way
\footnote{Subcurvature fluctuations cannot by themselves give rise to
  inflating islands since their size is smaller than the curvature
  scale.}.  Note that the r.m.s. amplitude for the $l=0$
supercurvature mode is a factor $\gamma^{-1/2}$ larger than the
amplitude for $l>0$ modes (recall that $\gamma\ll 1$). Hence, even if
the r.m.s. of $\phi$ is far below $M_p$, there is a certain
probability for $\phi$ to reach $M_p$ in a certain region near the
origin. The spherically symmetric mode is the one that is most likely
to contribute to this possibility. Even though there is a small
probability for this to happen, it is clear that only those rare
regions with $\phi\sim M_p$ will undergo a second stage of inflation;
so they will be the only ones that matter.  The value of the field on
those inflating islands will have the radial dependence of the $l=0$
mode, which decays as $\exp(-\gamma r/2)$ at large distances, $r\gg 1$.

Let us now discuss the more general models where the slow-roll field
has a small mass or it is massless outside the bubble. In this case,
one may ask what happens when a bubble nucleates in a place where the
slow-roll field already had a large (classical) value. This may occur,
for instance, if the whole universe was created at a large value of
$\phi$, and at the time when the bubble nucleates $\phi$ is still
rolling down from large values.  This possibility would in principle
be relevant for bubbles nucleated at early times, and is the one
considered in~\cite{GM}. However, as time goes by, the initially large
classical value of the slow-roll field in the false vacuum will
decrease, and all that will remain are the quantum fluctuations which
should be well described by the $O(4,1)$ or de Sitter-invariant
quantum state.

Occasionally, fluctuations of the slow-roll field in the false vacuum
may create a localized region with a higher value of the field. The
nucleation of a bubble on top of one of these regions will not be very
different from the case discussed in the previous paragraph.  Whether
the bubble nucleates on one of these high peaks or not, the field
outside the bubble will continue to fluctuate, and the bubble walls
will from time to time bump into regions with a higher value of the
field, as discussed in Section V.B.  Hence, also in this case, there
will be an ensemble of inflating regions with some distribution inside
the bubble. From a formal point of view, notice that the appearance of
the bubble has selected a point in spacetime, thus breaking the
$O(4,1)$ invariance, but otherwise respects a residual $O(3,1)$
symmetry. Therefore it seems reasonable to expect that, at least in a
statistical sense, the field inside the bubble will be well described
by the $O(3,1)$ invariant Gaussian distribution, corresponding to the
quantum state we have studied.  Just as in the case of the
supernatural model, here we also expect that the high peaks which lead
to inflating islands will have spherical symmetry, and the value of
the field on those islands will have the radial dependence of the
$l=0$ mode, which decays as $\exp(-\gamma r/2)$ at large distances
from the center of the island.

The co-moving size of the inflating islands is $r\sim \gamma^{-1}\gg
1$.  Since the volume on the hyperboloid grows exponentially with the
distance to the origin, as $\sinh^2 r dr$, a typical observer in a
quasi-open universe is most likely to be at $r\gg 1$, where the scalar
field behaves as $\phi\propto \exp(-\gamma r/2)$. Up to exponentially
small corrections, this is the same radial dependence that was
considered in~\cite{GM}. In that case, the fields were uncoupled and
$\gamma=2 m_F^2/3 H_F^2$. The arguments used in~\cite{GM} to estimate
the temperature anisotropies measured by a typical observer can be
directly applied to the models discussed here. Changing from the
coordinates $(r,\theta,\phi)$ to a new set $(r',\phi',\theta')$ such
that the point $r=r_0, \theta=0$ (with $r_0\gg 1$) is the new origin
of coordinates, one finds that the perturbation of the field $\phi$
around $r'=0$ can be described as~\cite{GM} $\delta \phi = \phi_0(t)
(\gamma/2) \ln f$, where $f\equiv (\cosh r' + \sinh r' cos \theta')$
and $\phi_0$ is the value of the field at the point $r'=0$.  The
corresponding gauge invariant potential at horizon crossing is 
\begin{equation}
\Phi\approx {3\over 5} \left.{H_T \delta \phi \over \dot \phi_0}
\right|_{t\approx H_T^{-1}}=
{3\over5}{H_T^2\over m_T^2}{\gamma\over2} \ln f(r',\theta').
\label{anis}
\end{equation}
The effect on the microwave background temperature fluctuations can be
computed by integrating the Sachs-Wolfe effect along the line of
sight~\cite{SW}. The dominant effect is in the
quadrupole~\cite{GM}, and it is of order
\begin{equation}
\left.{\delta T \over T}\right|_{l=2}\sim 10^{-1}
{H_T^2\over m_T^2} {\gamma\over 2}(1-\Omega_0).
\end{equation}
This is just a very rough order of magnitude estimate,
which works well for $\Omega_0 \gtrsim 0.3$.
A more detailed study of the power spectrum of temperature
anisotropies will be presented elsewhere~\cite{GBGM}.

Note that if the universe is sufficiently flat, the factor
$(1-\Omega_0)$ may completely erase the effect. Otherwise, for a
universe with appreciable curvature, we obtain a constraint on
$\gamma$,
\begin{equation}
\gamma\lesssim {2 \over (1-\Omega_0)}{m_T^2\over H_T^2}\,10^{-4}.
\label{semiclassical}
\end{equation}
where we have used $\delta T/T \lesssim 10^{-5}$, from the requirement
that this effect does not dominate the temperature anisotropies on
large scales, as seen by COBE~\cite{COBE}. Since $H_T$ is larger
than $m_T$, it is clear that $\gamma$ has to be very small in order to
avoid large temperature fluctuations.

In the supernatural model, $\gamma \sim R_0^2 m^2/2$, this constraint
implies 
$$
R_0 \lesssim 2\times 10^{-2} (1-\Omega_0)^{-1/2} H_T^{-1}
$$
which is not difficult to accomodate. Note that the size of the
bubble $R_0$ is necessarily less than the Hubble radius in the false
vacuum and hence can easily be much less than the Hubble radius in the
true vacuum. However, in the decoupled model discussed in~\cite{GM}
the constraint (\ref{semiclassical}) forces $H_T$ to be quite large,
and this causes a problem of large quantum fluctuations in the
supercurvature modes~\cite{misao}.

In the class of models (\ref{general}), the size of $\gamma$ is
determined by Eq. (\ref{gamma}). Clearly, the effect can be made small
by choosing parameters such that $m_F$ and the size of the bubble are
sufficiently small. This is not always straightforward to implement.
For instance, the ``hybrid'' open inflation model considered in
Ref.~\cite{GBLhybrid} turns out to be quasi-open and suffers from too
large semiclassical anisotropies. However, it is possible to write an
open hybrid model, with a massless inflaton in the false vacuum, that
satisfies the constraints~\cite{GBGM}.

\subsection{Supercurvature anisotropies}

In the previous subsection we have considered the case where the $l=0$
mode was ``oversized'', meaning that it took an amplitude much larger
than its expected r.m.s.  Because of this, an observer far from the
center of the inflating region would see the anisotropy (\ref{anis}).
In this section we shall estimate the anisotropies caused by the $l>0$
supercurvature modes. Here we are not thinking that these higher modes
are ``oversized''; they simply take random values of the order of
their r.m.s.  For simplicity, we shall consider an observer located at
$r=0$, but the effect should not be much different for an observer
located elsewhere.

The size of CMB anisotropies caused by the $l>0$ supercurvature 
modes has been estimated in~\cite{super,misao}. For the class of 
models (\ref{general}), where the supercurvature mode is normalized as
in (\ref{usual}), the quadrupole CMB anisotropies are of order
\begin{equation}
\left.{\delta T\over T}\right|^{sup}_{l=2} \sim (1-\Omega_0) {H_F\over H_T}
\left.{\delta T\over T}\right|^{sub}_{l=2}\,. \label{superanisg} 
\end{equation}
Here $(\delta T/ T)^{sub}$ are the temperature anisotropies caused by
the subcurvature modes (with $p^2>0$). The supercurvatre effect
decreases very fast with multipole number, basically as
$(1-\Omega_0)^{l/2}$.  If the fluctuations we observe in the CMB are due
to inflation, then we need $(\delta T/ T)^{sub} \sim 10^{-5}$, and
from (\ref{superanisg}) we have that $H_F$ cannot be too much larger
than $H_T$, unless the universe is almost flat.

For the supernatural model, the supercurvature mode (\ref{rmsdeg}) has
a normalization $2/(H_F R_0)$ times larger than its counterpart
(\ref{usual}) [we are ignoring the mild enhancement due to the factor
$(v/\sigma_b)$]. Hence, the analog of (\ref{superanisg}) is
\begin{equation}
\left.{\delta T\over T}\right|^{sup}_{l=2} \sim (1-\Omega_0) {2\over H_T R_0}
\left.{\delta T\over T}\right|^{sub}_{l=2}\,. \label{superaniss}
\end{equation}
Therefore we need $R_0 \gtrsim H_T^{-1}$.  Since $R_0$ has to be
necessarily smaller than $H_F^{-1}$, we have a two-fold restriction.
On one hand, $R_0 \sim H_F$, and on the other $H_T\sim H_F$. Thus,
it seems fair to say that the model is not as natural as it was
thought to be~\cite{LM}: the difference in energy density between the 
true and the false vacuum cannot span many orders of magnitude.
The reason is the following: In spite of the fact that the field is
massive in the false vacuum, a supercurvature mode exists. Its
normalization is not proportional to $H_F$ as in the usual case
(\ref{general}), but to $R_0^{-1}$, which is even larger.  The effect
can be thought as the excitation of the pseudo-Goldstone modes due to
the acceleration of the domain wall ``boundary''.  The model may still 
be viable in a certain range of parameters. Determining this range 
requires detailed analysis, which is left for future research~\cite{GBGM}. 

\section{Conclusions}

Open inflation is an appealing way of reconciling an infinite open
universe with the inflationary paradigm. In this scenario, a
symmetric bubble nucleates in de Sitter space, and its interior
undergoes a second stage of slow-roll inflation to almost flatness.
Single-field models of open inflation can in principle be constructed,
but it does not seem possible to do so without a certain amount of
fine tuning~\cite{LM}. The basic problem is that there is a hierarchy
between the large mass needed for successful tunneling and the small
mass required for successful slow-roll. For that reason, it seems
natural to consider two-field models of open inflation~\cite{LM} where
one field does the tunneling and the other drives slow-roll inflation
inside the bubble.

In this paper we show that a large class of two-field models of open
inflation do not lead to infinite open universes, as it was previous
thought, but to an ensemble of inflating islands of finite size. The
reason is that the quantum tunneling does not occur simultaneously
along both field directions, and the equal-time hypersurfaces in the
open universe are not synchronized with equal-density or fixed-field
hypersurfaces.  Technically, one finds that there are no $O(4)$
invariant instantons for the two-field system which would describe the
formation of a bubble with ``large'' values of the slow-roll field in
its interior.  Large values of the inflaton field, needed for the
second period of inflation inside the bubble, only arise as localized
fluctuations. The interior of each nucleated bubble will contain an
infinite number of such inflating regions, giving rise to a rather
unexpected form of the large scale structure of the universe in these
models. 

The picture is the following. Right after the bubble has nucleated
there will be, on the $t=const.$ hypersurfaces inside the bubble, a
certain density of occasional large fluctuations of the slow-roll
field that lead to inflating islands. This fluctuations are caused by
modes whose wavelength is larger than the curvature scale. Denoting by
$\gamma$ the eigenvalue of the Laplacian on the unit Hyperboloid (with
$\gamma<1$), the co-moving size of the inflating islands is given by
$d\sim\gamma^{-1}$ [The parameter $\gamma$ can be determined in terms
of the parameters of the model, see Eqs. (\ref{gamma1}) and
(\ref{gamma}), and it is important in deriving observational
constraints]. Each one of the inflating islands will be a quasi-open
universe.  Since the volume of the hyperboloid is infinite, inflating
islands with all possible values of the field at their center will be
realized inside of a single bubble. We may happen to live in one of
those patches where the universe appears to be open. The fact that the
inflating regions are finite gives rise to classical anisotropies like
those discussed in Ref.~\cite{GM}.

\

In particular, we have studied the supernatural model introduced by
Linde and Mezhlumian~\cite{LM}. We have shown that in spite of the
large mass of the inflaton field in the false vaccuum, there is a
supercurvature mode. Its amplitude is proportional to $R_0^{-1}$,
rather than the usual $H_F$. Here $R_0$ is the radius of the bubble at
the time of nucleation and $H_F$ is the Hubble rate in the false
vacuum.  Since $R_0^{-1}>H_F$, this effect is quite important. In
order to make the model compatible with observations, it is required
that the energy density in the false vacuum should not be much larger
than in the true vacuum. This means that $H_F/H_T$ cannot span many
orders of magnitude, as it was previously believed~\cite{LM}. The
supercurvature mode can be understood as the pseudo-Goldstone mode
associated with the choice of a tunneling direction in field space.
Combining the supercurvature anisotropies with the classical ones we
find that the range of $\Omega_0$ will also be restricted. Detailed
analysis is required in order to determine the range of parameters in
which the model may still be viable~\cite{GBGM}.

For the more general class of models (\ref{general}), the size of the
inflating islands can be chosen to be comfortably large by an
appropriate choice of parameters. In this way, the classical
anisotropy will be unobservably small. By order of magnitude, the
constraint is given by Eq. (\ref{semiclassical}), where $\gamma$ is
given in Eq. (\ref{gamma}). The constraint will be satisfied if the
mass of the slow-roll field is sufficiently small in the false vacuum
{\em and} $R_0$ is much smaller than $H_F^{-1}$. In a future
publication~\cite{GBGM} we will give more precise constraints from the
observed power spectrum of temperature anisotropies of the CMB.

Finally, there are some two-field models of open inflation, such as
the one introduced by Green and Liddle~\cite{induced} in the context
of induced gravity, which need not be affected in principle by the
classical anisotropies mentioned above. In these models, the value of
$\Omega_0$ is not variable; it is determined in terms of the
parameters in the potential.  It would be interesting to check whether
$O(4)$ symmetric instantons do indeed exist in this model.

\section*{Acknowledgements}

J.G.B. and J.G. thank Andrei Linde for very stimulating discussions.
J.G. also thanks Alex Vilenkin for very useful conversations, and the
Theory Division at CERN for their hospitality during part of this
work. J.G. and X.M.  acknowledge financial support from CICYT
under contract AEN95-0882 and from European Project CI1-CT94-0004.

\appendix
\section{}

We will show in this Appendix that the two-point function on a
$t=const$ surface for the state $\varphi^{\gamma}$ dies off as
$e^{-\gamma d/2}$, where $d$ is the comoving distance between the
points. We will compute the two-point function ouside the lightcone,
and then continue it to the inside.

To compute the two-point function for $\varphi_2$, hereafter
$G_{\varphi_2}(x,x')$, we will use the fact that the modes ${\cal Y}_{plm}$
are properly normalized as Klein-Gordon modes of mass $p^2+1$ living in the
$\tau = const.$ (2+1) de Sitter hypersurfaces of the outside lightcone
metric. Thus we can define the fields
\begin{equation}
{\cal Y}_{p} = \sum_{lm} {\cal Y}_{plm}\, {}^{dS}a_{lm}^{(p)} + h.c,
\end{equation}
and the de Sitter invariant vaccuum $|0\rangle_{dS}^{(p)}$ anihilated
by $^{dS}a_{lm}^{(p)}$. Notice that $d\Omega_{dS}$ in (\ref{metric1})
corresponds to the line element of a closed coordenization of a (2+1)
de Sitter space, for which the two-point functions can be found in
Ref.~\cite{candelas}.

We can now write the two-point function $G_{\varphi_2}$ in terms of the 
two-point functions for ${\cal Y}_p$,
\begin{eqnarray}
G_{\varphi_2}(x,x') &=& \langle0|\varphi_2(x)\varphi_2(x')|0\rangle = 
a_E^{-2}\sum_{plm} F_p(\tau)\overline{F_p(\tau')} {\cal Y}_{plm}(x^i) 
\overline{{\cal Y}_{plm}(x'^i)} \nonumber\\ \label{twopointsum}
&=&a_E^{-2}\sum_p F_p(\tau)\overline{F_p(\tau')} 
{}_{dS}\langle0|{\cal Y}_p(x^i){\cal Y}_p(x'^i)|0\rangle_{dS},
\end{eqnarray}
where the sum over p has to be understood like a sum over the discrete
eigenvalues $p^2$ of the Schr{\"o}dinger equation (\ref{schrodinger})
and like an integration over its continuum spectrum $p^2>0$. On a
given $\tau=const$ hypersurface, the $\rho$ dependence of
$G_{\varphi_2}(x,x')$ will be given by $G_p(x^i,x'^i)$, weighted for
each $p$ by $a_E^{-2}||F_p||^2$. The two-point function $G_p$ can be
found in~\cite{candelas} 
\begin{equation}
\label{twopoint}
G_p(\xi,\xi') = \frac{1}{(4\pi)^{3/2}}
\frac{\Gamma(1-i p)\Gamma(1+i p)}{\Gamma(3/2)} 
F\left(1+i p,1-i p;\frac{3}{2};\frac{1+Z}{2}\right),
\end{equation}
where $F$ is the hypergeometric function and $Z$ is the scalar product of the
position vectors at points $x^i$ and $x'^i$ in the embedding (3+1) Mikowski
space,
\begin{equation}
\label{scalarprod}
Z(x^i,x'^i) = \xi^\mu(x^i) \xi_\mu(x'^i) = 
cos \tilde\gamma \cosh \rho \cosh \rho'-
\sinh\rho \sinh\rho'.
\end{equation}
Here $\tilde\gamma$ is the angle on the 2-sphere between the two
points. We recall that for the lowest discrete eigenmode to first
order in the shift $\gamma$, $i p = 1 - \gamma/2$, so we will denote
by $G^\gamma$ the two-point function for this eigenmode.

Now we have to analytically continue (\ref{twopoint}) to the inside of the
lightcone by means of (\ref{analytical}). This amounts only to analytically
continuing the scalar product $Z$,
\begin{equation}
  Z(x^i,x'^i) \rightarrow - cos \tilde\gamma \sinh r \sinh r' + \cosh r
  \cosh r'.
\end{equation}

Taking $r'=0$ and $r =d$, so $Z = \cosh d$, and using eq. (9.131.1) in
Ref.~\cite{formules}, we find that inside the light-cone the two-point
function between points separated a comoving distance $d$ can be
written as
 \begin{equation}
\label{twopointinside}
G^\gamma(d) = \frac{1}{(4\pi)^{3/2}}
\frac{\Gamma(2-\gamma/2)\Gamma(\gamma/2)}{\Gamma(3/2)} \left(\frac{1-\cosh
    d}{2}\right)^{-\gamma/2}
F\left(\frac{\gamma}{2},\frac{\gamma-1}{2};\frac{3}{2};
\frac{1+\cosh d}{\cosh d-1}\right).
\end{equation}
As $d \rightarrow \infty$, the hypergeometric function in
(\ref{twopointinside}) tends to a constant, and the assymptotic behaviour of
$G^\gamma$ is given by
\begin{equation}
  G^\gamma(d)\rightarrow \frac{1}{(4\pi)^{3/2}}
  \frac{\Gamma(\gamma/2)\Gamma(2-\gamma)}{\Gamma(3/2 -
    \gamma/2)}\left(\frac{-1}{4}\right)^{-\gamma/2}\,e^{-\gamma d/2},
\end{equation}
which dies off exponentially with $d$.

Here, we have only computed the first term in the sum (\ref{twopointsum}). The
terms with $p^2>0$ decay as $e^{-r/2}$, and hence they are subdominant at
large distance.

\section{}

To compute $(\Delta\varphi_2^\gamma)^2$ for large $r$, we will need the
asymptotic expressions for the hyperbolic harmonics ${\cal Y}_{\Lambda,lm}$
for $r\gg \gamma^{-1}$. The Legendre functions are given by (see e.g. 
Ref.~\cite{AS})
\begin{equation}
\label{legzero}
P^{-1/2}_{\nu-1/2}(\cosh r) = \sqrt{2\over\pi}
{\sinh \nu r\over\nu\sqrt{\sinh r}},
\end{equation}
{}From equation (\ref{hyper}),
the supercurvature mode ${\cal Y}_{\Lambda,00}$ is given by 
\begin{equation}
{\cal Y}_{\Lambda,00} = {1\over\pi}\left[
\frac{\Gamma(1-\Lambda)\Gamma(1+\Lambda)}{4}\right]^{1/2}
\frac{\sinh \Lambda r}{\Lambda\sinh r},
\end{equation}
where $\Lambda = 1-\gamma/2$. For large $r$, we have
\begin{equation}
  {\cal Y}_{(1-\gamma/2),00} \to {e^{-r\gamma/2}\over\pi\sqrt{2\gamma}}
  [1 + O(\gamma)].
\end{equation}
For $l>0$, we can express $P^{-l-1/2}_\nu$ in terms of $P^{-1/2}_{\nu+k}$
using the recursion formula
\begin{equation}
\label{recursion}
P^{\mu}_\nu (z) = \frac{1}{2\nu + 1}\frac{1}{\sinh r}
\left(P^{\mu-1}_{\nu+1}(z) - P^{\mu-1}_{\nu-1}(z)\right).
\end{equation}
The Legendre function $P_\nu^{-l-1/2}$ acquires then the form
\begin{equation}
\label{sum}
P^{-l-1/2}_\nu = \frac{1}{\sinh^l r}\sum_{k=-l}^l C_k(\nu)
P^{-1/2}_{\nu+k},
\end{equation}
where $C_k(\nu)$ are some functions depending on $\nu$. In fact, for large
$r$, we do not need to compute all $C_k(\nu)$. We have to take into account
that for a supercurvature mode, $P^{-1/2}_\nu(\cosh r)$ behaves for large $r$
as $e^{|\nu|r}$ (as can be seen from (\ref{legzero})). Thus, for the
supercurvature mode $\nu = (1-\gamma)/2$, the term $k=l$ in (\ref{sum}) grows
exponentially faster than the rest of terms in the sum, so the main
contribution for large $r$ will be given by this term. The coefficienty
$C_l(\nu)$ can be easily read from (\ref{recursion}):
\begin{equation}
\label{important}
C_l(\nu) = \frac{\Gamma(\nu+1/2)}{2^l\Gamma(\nu+1/2+l)}.
\end{equation}
For large $r$, using (\ref{sum}), (\ref{important}) and (\ref{hyper}),
we obtain
\begin{equation}
  {\cal Y}_{(1-\gamma/2),lm} \to
  \left[\frac{\Gamma(l+\gamma/2)}{\pi(1-\gamma/2+l)
      (l-\gamma/2)\Gamma(l-\gamma/2)}\right]^{1/2}\Gamma(1-\gamma/2)
  e^{-\gamma r/2} Y_{lm}(\Omega).
\end{equation}
Finally, using $\lim_{x\to\infty}\Gamma(x+a)/\Gamma(x)= x^a$,  
we can write ${\cal Y}_{(1-\gamma/2),lm}$ for large $l$, to order
$\gamma$, as
\begin{equation}
  {\cal Y}_{(1-\gamma/2),lm} \approx \frac{l^{\gamma/2-1}}{\sqrt \pi} 
 e^{-r\gamma/2} Y_{lm}(\Omega)[1 + O(\gamma)].
\end{equation}

Using the results derived above, we can compute the amplitude of the $l=0$
mode near the origin,
\begin{equation}
  \left.(\varphi_{l=0}^\gamma)^2\right|_{r=0} = \left.({\cal N}\sigma_0)^2
    {\cal Y}_{\Lambda,00} \overline{{\cal Y}_{\Lambda,00}}\right|_{r=0}
  \approx \frac{1}{\pi^2R_0^2\gamma}
  \left(\frac{\sigma_0}{\sigma_b}\right)^2[1+O(\gamma)],
\end{equation}
and for large $r$,
\begin{equation}
(\varphi_{l=0}^\gamma)^2 \to \frac{1}{\pi^2R_0^2\gamma}
\left(\frac{\sigma_0}{\sigma_b}\right)^2 e^{-r\gamma} =  
 \left.(\varphi_{l=0}^\gamma)^2\right|_{r=0} e^{-r\gamma}
\end{equation}
As we can see, the amplitude of the mode $l=0$ decays exponentially
for $r\gg \gamma^{-1}$. Taking into account that we have choosen an
$O(3,1)$ symmetric vaccuum, this decrease in amplitude for large $r$
must be compensated by the joint contribution of the $l>0$ modes,
smeared over a suitable length scale, in such a way that the r.m.s
fluctuations of the field are independent of $r$. Let us check it. We
need to compute
\begin{eqnarray}
  (\Delta \varphi_2^\gamma)^2 &\equiv& \sum_{l=1}^{l_*}\sum_{m={-l}}^{l}
  \varphi^\gamma_{i\Lambda ,lm} \overline{\varphi^\gamma_{i\Lambda ,lm}} \to
  ({\cal N}\sigma_0)^2\sum_{l=1}^{l=l_*} \frac{(2 l
    +1)}{4\pi}\frac{l^{\gamma-2}}{\pi} e^{-\gamma r} \nonumber\\ &\approx&
  ({\cal N} \sigma_0)^2 \frac{e^{-\gamma r}}{2 \pi^2}\int^{l_*}_0
  l^{\gamma-1}\,dl \approx ({\cal N} \sigma_0)^2 \frac{e^{-\gamma r}}{2
    \pi^2}\frac{l_*^\gamma}{\gamma},
\end{eqnarray}
where $l_*$ is a certain cutoff, which has to grow as we move away
from the origin to include more and more modes in the sum. If we smear
the field over a fixed comoving lenght~$\xi$, realizing that the
wavelength of the $l$-th multipole is proportional to $(\sinh r)/l$, we
can take $l_* = \sinh r/\xi$.  Finally, we obtain
\begin{equation}
(\Delta \varphi_2^\gamma)^2 \to \frac{1}{\pi^2 R_0^2\gamma}
\left(\frac{\sigma_0}{\sigma_b}^2\right)(2\xi)^{-\gamma} \approx
\left.(\varphi_{l=0}^\gamma)^2\right|_{r=0} (1 - \gamma\ln 2\xi).
\end{equation}
As we can see, as long as $|\ln 2\xi|\ll \gamma^{-1}$, the added
contribution of the relevant modes is the same as the one given by the
$l=0$ mode near the origin.

\section{}

In the thin wall approximation, neglecting gravitational backreaction, the
background geometry is found~\cite{thingrav} to be described by two de Sitter
pieces with different Hubble constant glued together at some $\eta_W$. The
scale factor is given by
\begin{equation}
  \label{scalefactor}
  a_E(\eta) = a_F(\eta)\theta(\eta-\eta_W) + a_T(\eta)\theta(\eta_W-\eta),
\end{equation}
where $a_F$ and $a_T$ are the scale factors in the false and in the true
vaccuum,
\begin{eqnarray*} 
  a_F(\eta) &=& \frac{1}{H_F \cosh\eta} \nonumber\\
  a_T(\eta) &=& \frac{1}{H_T \cosh (\eta - \delta)}
\end{eqnarray*}
Continuity of $a_E$ at the wall implies
\begin{equation}
  \label{glued}
  a(\eta_W) = \frac{1}{H_F \cosh \eta_W} = \frac{1}{H_T \cosh (\eta_W - \delta)}
  = R_0,
\end{equation}
where $R_0$ is the radius of the wall, and $\delta$ is given by
\begin{equation}
  \label{delta}
  e^{\delta} = \frac{(1+\sqrt{1-H_T^2R_0^2})(1-\sqrt{1-H_F^2 R_0^2})} {H_F H_T
    R_0^2}.
\end{equation}
To complete the description, we need to know the value of $R_0$. It can be
found in Ref.~\cite{thingrav}:
\begin{equation}
R_0={\kappa S_1\over \sqrt{(H_F^2-H_T^2+(\kappa S_1/2)^2)^2+
\kappa^2 H_T^2 S_1^2}},
\end{equation}
where $\kappa=8\pi G$ and $S_1$ is the wall tension. 

We want to find the lowest eigenvalue of the Schr{\"o}dinger equation
(\ref{schrodinger}) in the background given above. The effective potential is
given in this case by
\begin{equation}
  {\cal U} = a_E^2 [m^2 + g\sigma_0^2 - 2 (H_F^2 \,\theta(\eta-\eta_W) +
  H_T^2 \,\theta(\eta_W-\eta))] + ({\cal H}_F-{\cal H}_T)\delta(\eta-\eta_W),
\end{equation}
where ${\cal H}_F = a'_F/a_F$ and similarly for ${\cal H}_T$.

We will take a perturbative approach. We will divide the effective potential
${\cal U}$ in an unperturbed one, ${\cal U}_0$, and in a small perturbation
part, $\lambda {\cal U}_1$:
\begin{eqnarray}
  {\cal U}_0 &=& -2 a_F H_F^2 \\
  \lambda{\cal U}_1 &=& a_E^2\left(m^2 + g\sigma_0^2 -2 H_T^2\theta(\eta_W-\eta)\right)
  + ({\cal H}_F-{\cal H}_T)\delta(\eta-\eta_W)\nonumber\\& &+2 H_F^2 a_F^2
  \theta(\eta_W-\eta).
\end{eqnarray}
The unperturbed ${\cal U}_0$ corresponds to the effective potential of
a massless scalar field in de Sitter space, which has as a ground
state a supercurvature mode with energy $p_0^2=-1$ and 
wavefunction~\cite{super}
\begin{equation}
F_{-1} = \frac{H_F}{\sqrt 2}a_F(\eta).
\end{equation}
To first order in perturbation theory, the shift of the energy $p^2_0=-1$ is
given by 
\begin{equation}
  \gamma = \langle-1|\lambda {\cal U}_1|-1\rangle = 
  {2\over 3}{m_F^2\over H_F^2}+
  {H_F^2 R_0^4\over 8}(m_T^2 -m_F^2),
\end{equation} 
where $m_F$ is the efective mass of the slow-roll field in the false vaccuum,
and $m_T$ the effective mass in the true vaccuum. In this case, $m_F^2=m^2$ and
$m_T^2=m^2+g v^2$.

\end{document}